\documentclass[prints]{aa}
\usepackage[authoryear]{natbib}
\bibpunct[, ]{(}{)}{;}{a}{}{,}
\usepackage{graphicx}

\newcommand{\feh}{\mbox{[Fe/H]}}
\newcommand{\zh}{\mbox{[Z/H]}}
\newcommand{\afe}{\mbox{[$\alpha$/Fe]}}

\begin{document}

\title{Globular cluster content and evolutionary history of NGC~147}
\titlerunning{Globular Clusters in NGC147}

\author{Margarita Sharina \inst{1,2} \and
Emmanuel Davoust \inst{2}}
\offprints{M. Sharina}
\institute{Special Astrophysical Observatory, Russian Academy
of Sciences, N. Arkhyz, KChR, 369167, Russia\\
\and Laboratoire d'Astrophysique de Toulouse-Tarbes, Universit\'e de Toulouse, CNRS, 14 avenue E.~Belin, F-31400
Toulouse, France\\}
\date{Received:  November 2008}

\abstract
{Globular clusters are representative of the oldest stellar populations.
It is thus essential to have a complete census of these systems in dwarf galaxies,
from which more massive galaxies are progressively formed in the hierarchical scenario.}
{We present the results of spectroscopic observations of eight
globular cluster candidates in  NGC~147, a satellite dwarf elliptical galaxy of M31.
Our goal is to make a complete inventory of the globular cluster system of
this galaxy, determine the properties of their stellar populations, and 
compare these properties with those of systems of globular clusters in other dwarf galaxies.}
{The candidates were identified on Canada-France-Hawaii telescope photographic
plates. Medium resolution spectra were obtained with the SCORPIO spectrograph
at the prime focus of the 6m telescope of the Russian Academy of Sciences.
They were analyzed using predictions of stellar population synthesis models.}
{We were able to confirm the nature of all eight candidates,
three of which (GC5, GC7, and GC10) are indeed globular clusters, and
to estimate evolutionary parameters for the two brightest ones and for Hodge~II.
The bright clusters GC5 and GC7 appear to have metallicities (\zh$\sim -1.5 \div -1.8$)
that are lower than the oldest stars in the galaxy.
The fainter GC Hodge~II has a metallicity \zh$=-1.1$, 
similar to that of the oldest stars in the galaxy.
The clusters GC5 and GC7 have low alpha-element abundance ratios.
The mean age of the globular clusters in NGC~147 is 9$\pm 1$ Gyr.
We also measured the radial velocities of Hodge~II and IV, and derived
a mass of NGC~147 in good
agreement with the value from the literature. The frequency, $S_n =6.4$, and mass fraction, $T=14$
of globular clusters in NGC~147 appear to be higher than those for NGC~185 and 205.}
{Our results indicate that the bright clusters GC5, GC7, and Hodge~III formed 
in the main star-forming period $\sim$8-10 Gyr ago,
while the fainter clusters Hodge~I and II formed together with the second generation of field stars.
}


\maketitle

\section{Introduction}

 The relative frequency of dwarf ellipticals (dEs) and Sculptor-type spheroidal galaxies (dSph)
depends on the environment, and one can expect that the properties of their populations of globular 
clusters (GCs) also do.  In the Local Volume (LV), which spans a region of up to $\sim$10 Mpc from the Sun,
the most frequent types are low surface-brightness dwarf irregular and spheroidal galaxies.
In contrast, there are hundreds of dEs in the nearest rich clusters of galaxies Virgo and Fornax
(Binggeli et al. 1985; Ferguson, 1989).
\begin{figure*}[!t]
\caption{Reproduction of the CFHT B-band photographic plate. The field of view is $18 \arcmin$. 
The globular clusters and candidates are indicated.}
\label{ps:image}
\end{figure*}
\begin{figure*}[!t]
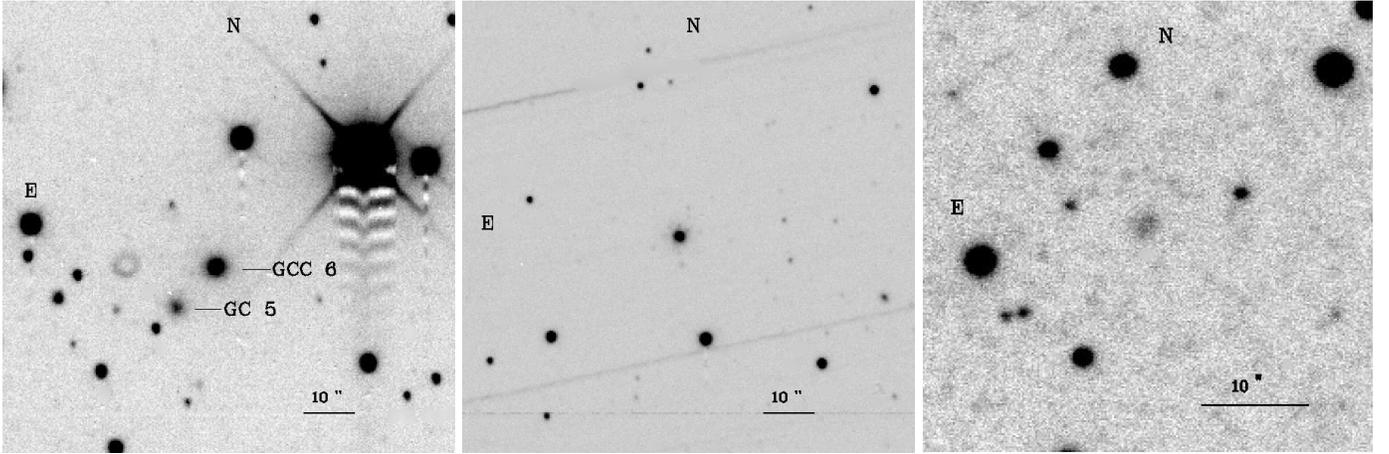

\includegraphics[width=6.0cm]{1306f2a.ps}
\includegraphics[width=6.0cm]{1306f2b.ps}
\includegraphics[width=6.0cm]{1306f2c.ps}
\caption{Pic-du-Midi and 6m telescope images of GC5, GC7, and GC10, taken in good atmospheric conditions (seeing $\sim 1 \arcsec$).}
\label{ps:images}
\end{figure*}

NGC~147, 185, 205, and M32 are the brightest early-type dwarf galaxies in the
LV (Mateo, 1998; van den Bergh, 2000; Karachentsev et al., 2004).
Their absolute V-magnitudes range from $-15.1$ for NGC 147 to $-16.7$ for M32, making them
more luminous than the Sculptor-type spheroidal galaxies (dSph), whose absolute V-magnitudes are
between $-8$ and $-13$. They also have higher central and mean surface brightnesses (see e.g. Karachentseva et al. 1985).
Their colors resemble those of old, gas--free stellar systems, such as the Galactic globular clusters.

The four satellite galaxies of M31 were first resolved into individual stars by Baade (1944a, b).
Their distances are well determined, and the properties of their stellar populations have been thoroughly studied
(e.g. Mould et al. 1983, Davidge, 1994, 2005; Lee et al., 1993; Grebel 2000,
McConnachie 2005; Butler \& Mart{\' i}nez-Delgado, 2005; Dolphin 2005).
NGC~147 is composed of old (90\%) and intermediate-age stellar populations (Mould et al. 1983; Davidge, 2005).
The percentage of intermediate-age stellar populations is higher and the metallicities and metallicity spreads 
are greater in the other dEs.

 NGC~185, 205, and M32 have prominent peculiarities in their structure and stellar content.
It has been argued that the gas-free M32 is a stripped spiral (Bekki et al., 2001; Graham, 2002).
NGC~185 and 205 contain neutral hydrogen and molecular gas, dust, and young stars
(Baade, 1944a; Baade, 1944b; Hodge, 1963; Lee et al., 1993,
Butler \& Mart{\' i}nez-Delgado, 2005; Young \&  Lo, 1997; Sage et al., 1998).
Both galaxies reveal bright nuclei with a young star-forming region in the first one, and
a relatively young star cluster in the other (see e.g. Da Costa \& Mould 1988; Young \&  Lo, 1997; Sharina et al., 2006b).

NGC 185 and 147 have similar photometric and structural properties (Mateo, 1998).
van den Bergh (1998) has shown that these galaxies form a binary system.
However, there are no indications of a tidal bridge between them (Battinelli \& Demers, 2004).
They are slightly nearer to our Galaxy than to M31, and are not supposed to have had close passages to M31.
The differences in the gas content
and properties of stellar populations between NGC 185 and 147 are not well understood.
The observed gas content in NGC~185 and 205 is explained by stellar mass return (Welch et al., 1998).
In contrast, in NGC~147 and M32 the mass of the interstellar medium is much lower than the one expected
from material ejected from evolved stars (Sage et al. 1998).

The most prominent properties of stellar populations, common to NGC~147, 185, and 205, are
i) sizable globular cluster systems,
where the majority of GCs are old and metal-poor (Da Costa \& Mould 1988, van den Bergh, 2000; Sharina et al., 2006b),
ii) the presence of intermediate-age stars (Davidge, 2005; Saha, Hoessel \& Mossman 1990),
planetary nebulae (Richer \& McCall, 1995; Corradi et al., 2005; Gon\c{c}alves et al. 2006),
 and iii) similar median metallicities of old stars, \feh$\sim -0.9$ dex
(Butler \& Mart{\'i}nez-Delgado, 2005; Davidge, 2005).
It is noteworthy that the stellar halo of M31 at galactocentric distances $R \ge 60$ kpc is
more metal-poor $\langle \feh \rangle = -1.26 \pm 0.10$ dex ($\sigma=0.72$) (Kalirai et al., 2006).

Recent observational results have demonstrated that the formation of GCs
occurs under high densities and pressures during the most powerful star forming events in the life of a galaxy.
This is why the determination of evolutionary parameters of GCs is important for understanding the star formation histories of individual galaxies
and the formation of cosmological structures in general.
Four globular clusters were found in NGC~147 by Baade (1944b), who indicated the existence of two GCs and
a "semistellar nucleus", and by Hodge (1976). Accurate positions for Hodge I--III were given by Ford et al. (1977).
Da Costa \& Mould (1988) were the first to obtain spectra for the brightest GCs in NGC147, 185, and 205, and estimated
their mean abundance to be considerably less than that
of the corresponding field halo stars. They concluded that the GCs were formed at
the earliest epochs of galactic formation along with the first stars.
Sharina et al. (2006b, hereafter SAP06) obtained high S/N ratio, medium-resolution spectra for two GCs in NGC~147, 6 GCs in NGC~185, and
5 GCs in NGC~205, and estimated evolutionary parameters for them and for diffuse galactic light regions near the centers of NGC~205 and NGC~185.

Our paper continues the aforementioned spectroscopic studies of GCs in NGC~147.
We found three additional globular clusters in NGC~147, obtained their spectra on the 6m telescope,
and estimated ages, metallicities and abundance ratios for the two brightest, GC5 and GC7, and for Hodge~II.
Additionally, we obtained radial velocities for Hodge~IV and GC10.
These new data, together with others gathered from the
literature, enabled us to propose a scenario for the evolutionary history of NGC~147.
Our paper is organised as follows. In Section 2 we describe our observations and the data reduction.
In Section 3 we explain our methods of age and metallicity determination. We discuss the obtained results in Section~4,
and formulate our conclusions in Section~5.
In the appendix we give a mass estimate for NGC~147 using radial velocities for the GCs obtained by us and
by SAP06, and present absorption-line indices measured in the spectra of GC5, GC7, and Hodge~II, as well as 
comparisons of our Lick index measurements with SSP model predictions.

\section{Observations and Data Reduction}

\subsection{Images}

This project started when non-stellar objects were discovered on photographic
plates of NGC 147 obtained at the Cassegrain focus of the 3.6m 
Canada-France-Hawaii telescope (CFHT).  These plates were in fact
the first ones ever taken at this focus of the telescope. 
Follow-up CCD imaging of the globular cluster candidates (GCCs) were obtained under good
seeing conditions at the F/10 Cassegrain focus of the 2m Bernard-Lyot telescope
of Pic-du-Midi Observatory. These preliminary imaging observations are
summarised in Table~\ref{tab:imalog}, together with those obtained during
the spectroscopic runs 2007, 2008.
The exact positions of the GCCs were determined by J.F. Le Campion at Bordeaux Observatory  :
a photographic plate of the field was obtained at the astrographic telescope, 
the positions of 64 stars in common with the CFHT plates were measured on all 
plates, as well as those of the GCCs, and their precise
coordinates determined using 47 fundamental stars. Finally, the coordinates
of the GCCs were derived with a precision of $ 0.55\arcsec$.
They are given in Table~\ref{tab:coord}, together with their magnitudes and colors
measured on the CCD images. The positions of the GCCs and of the four known
globular clusters (Hodge I to IV) are shown on Fig.~\ref{ps:image}, which is
a reproduction of the CFHT photographic plate in the B-band.
CCD images of the spectroscopically confirmed GCCs are presented in Fig.~\ref{ps:images}.

V and R magnitudes in the Johnson-Cousins photometric system were obtained for the GCCs 
using the pre-imaging observations, with typical errors $\sim 0.3$ mag. 
The main contributors to the errors are uncertainties of transformations between the
instrumental and standard systems using secondary standard stars and random photometric errors significant at low exposure
times and with old matrices.
 We could not use photometric standards to calibrate our measurements
because atmospheric conditions were unstable during the observations.
The magnitudes are gathered in Table~2 together with heliocentric radial velocities
and projected distances from the center of the galaxy.
Assuming a distance modulus $(m-M)_0=23.95$ mag (McConnachie et al. 2004)
and a Galactic extinction $A_V=0.58$ (Schlegel et al. 1998), the absolute V magnitudes of GC5, GC7, GC10, and Hodge~IV
are -5.7, -7.1, -4.0, and -4.7, respectively.  
The absolute magnitudes of Hodge~I, II, III were determined by Sharina \& Puzia (2009) on HST WFPC2 images.
The resulting mean absolute magnitude of GCs in NGC~147 is $M_V \sim -6.0$ mag.
This value is fainter than the one for globular cluster systems in giant galaxies, which is  $M_V = -7.5$,
fairly independent of the parent galaxy morphological type (Harris, 1991).

Full widths at half maxima (FWHM) for GC5, GC7, and GC10 approximately correspond
to half-light radii of 2, 1.5, and 3 pc, respectively,
if one convolves artificial two-dimensional frames calculated using
the Gaussian functions of corresponding FWHM with the observed stellar point-spread functions.
These values are typical of GCs in dwarf ellipticals in the Virgo cluster (Jord{\' a}n et al., 2005).
\begin{figure*}[!t]
\includegraphics[width=6.70cm,angle=-90]{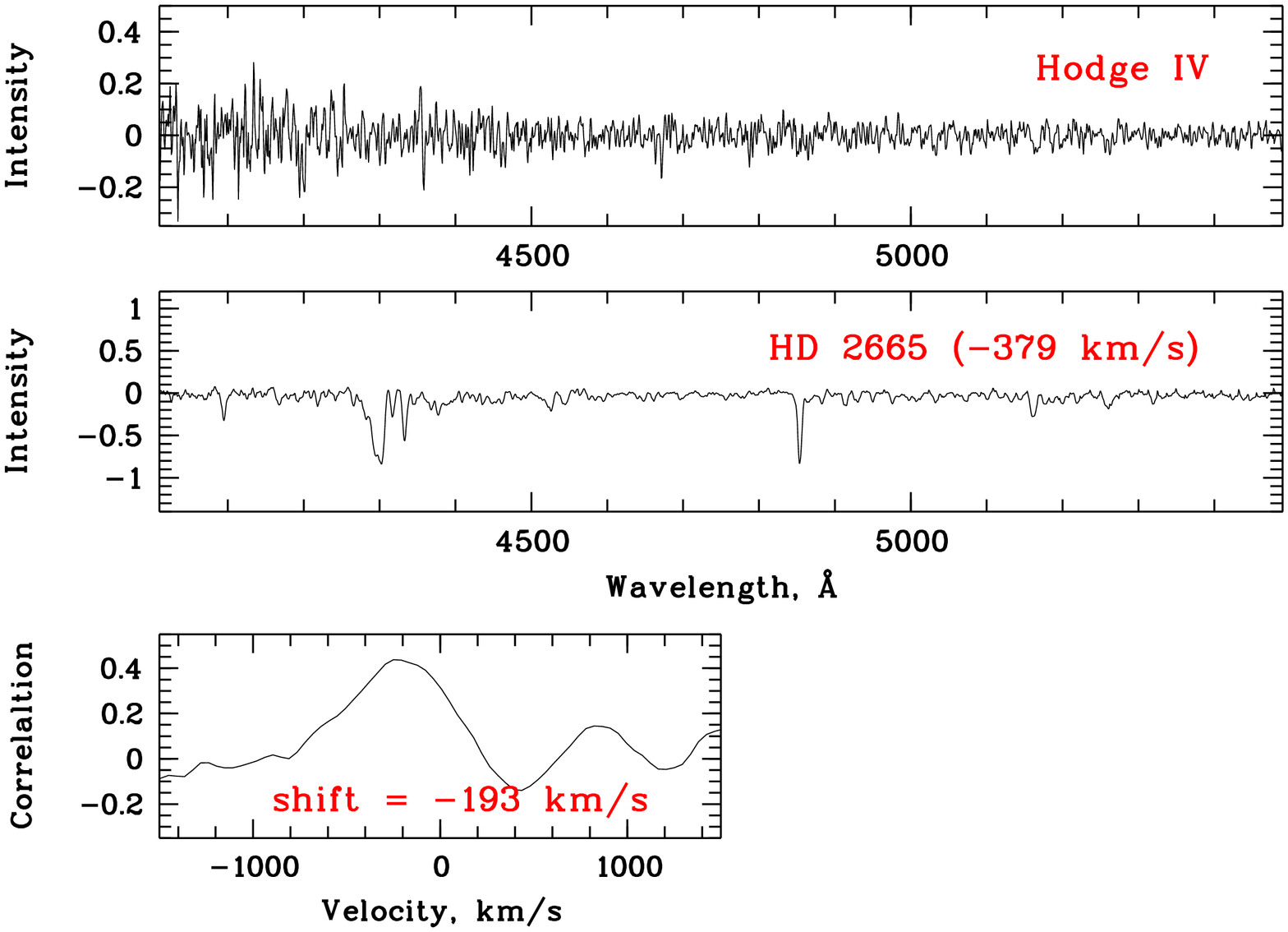}
\includegraphics[width=6.70cm,angle=-90]{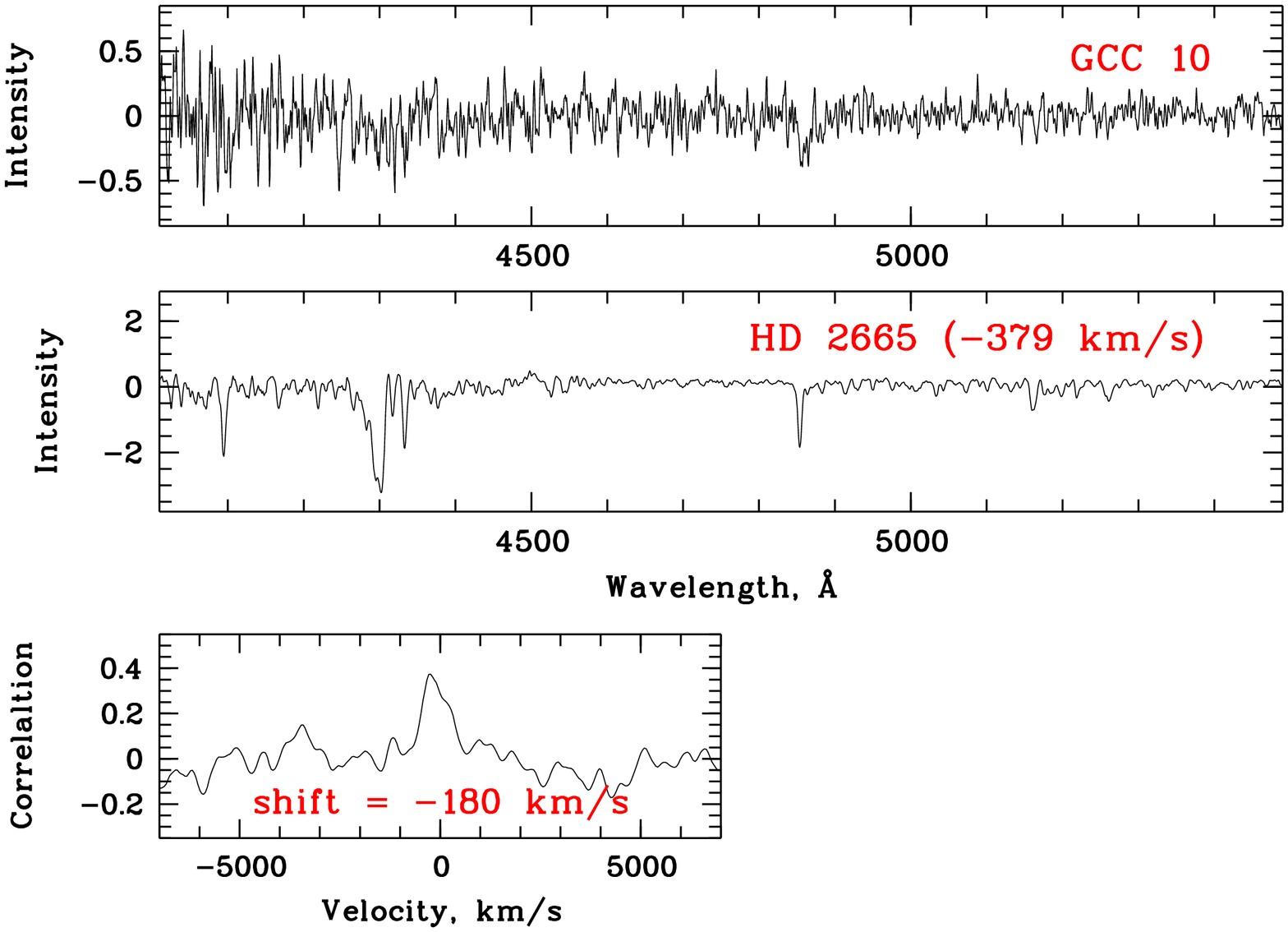}
\includegraphics[width=6.70cm,angle=-90]{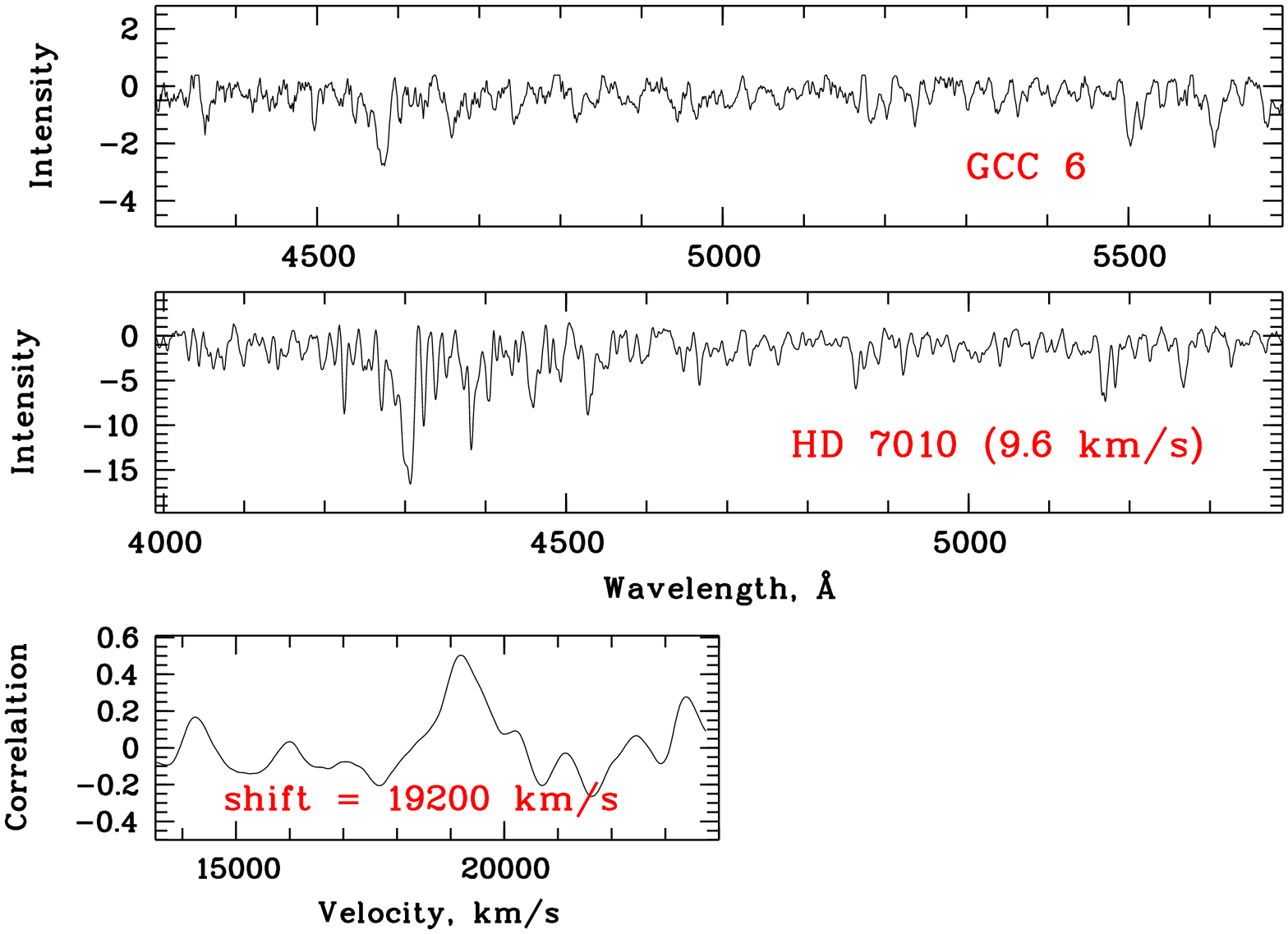}
\includegraphics[width=6.70cm,angle=-90]{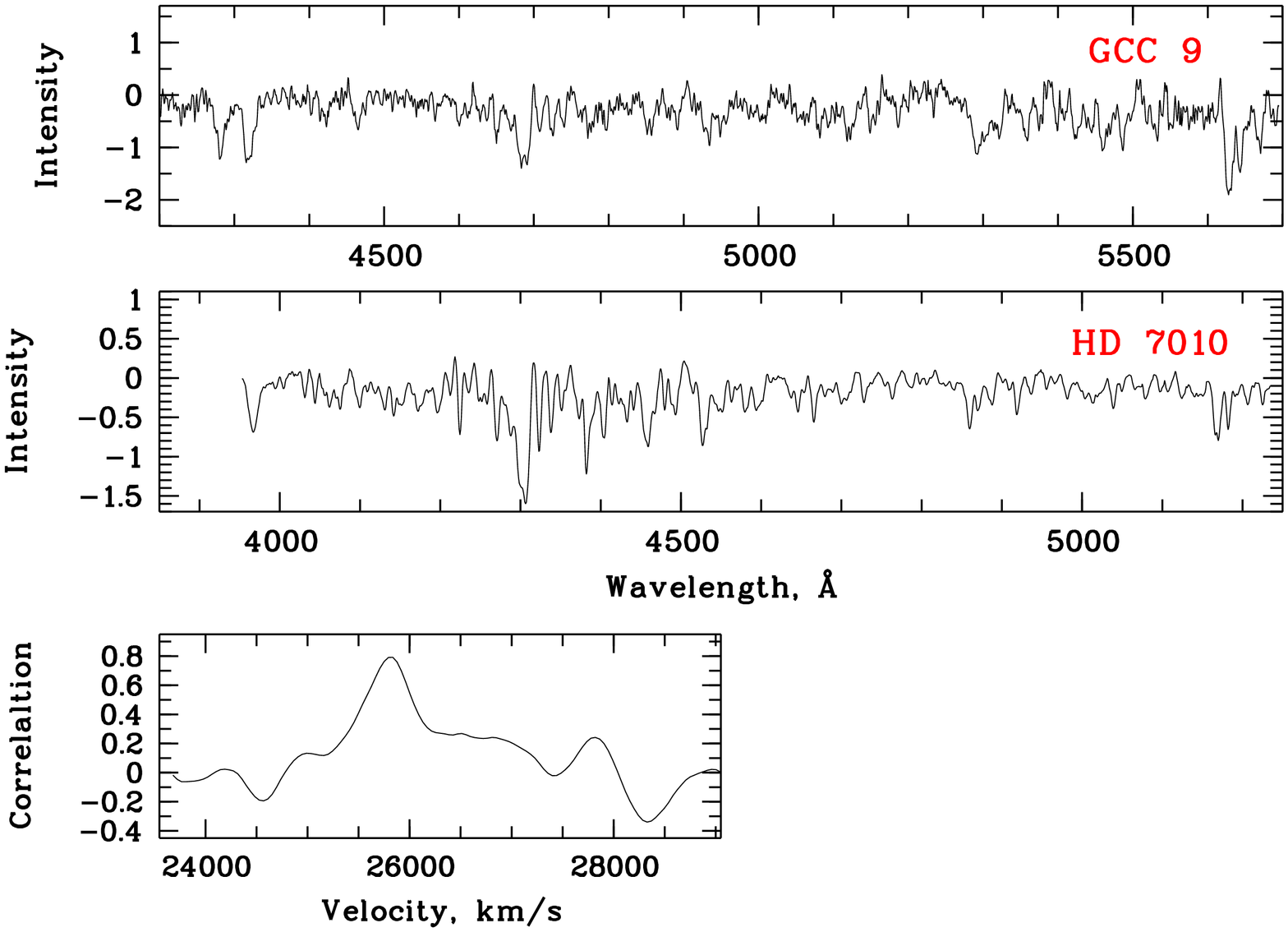}
\includegraphics[width=4.92cm,angle=-90]{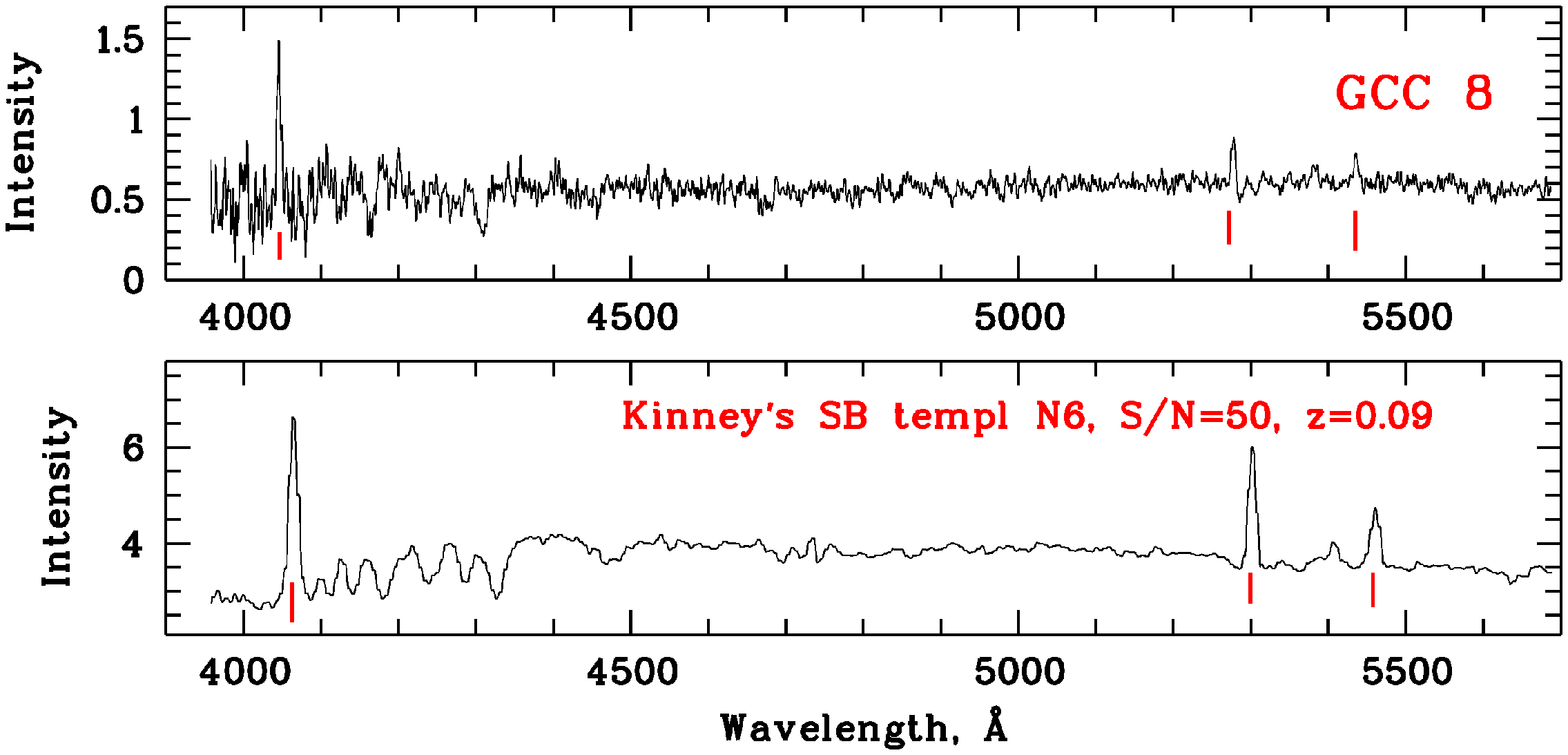}
\includegraphics[width=4.92cm,angle=-90]{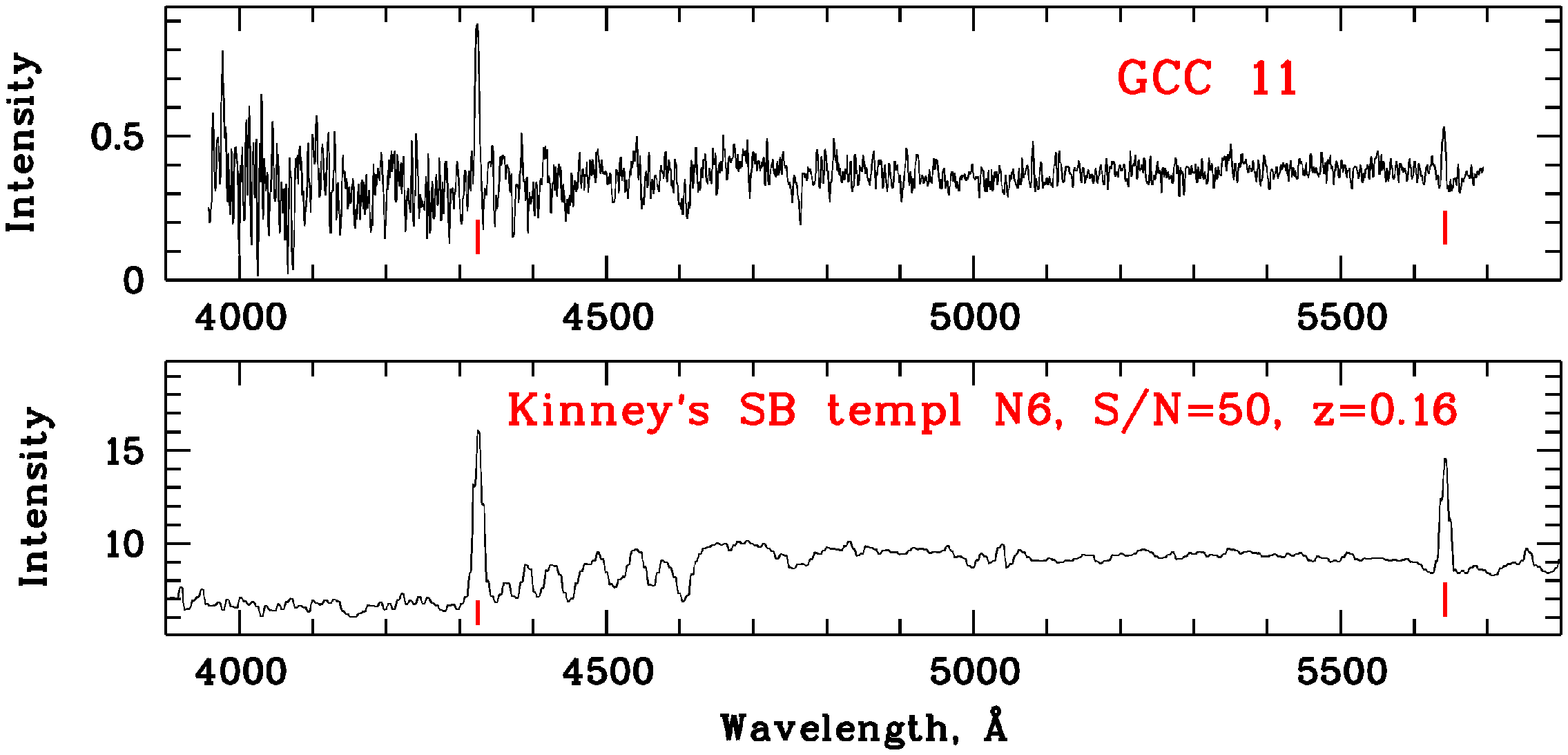}
\includegraphics[width=4.92cm,angle=-90]{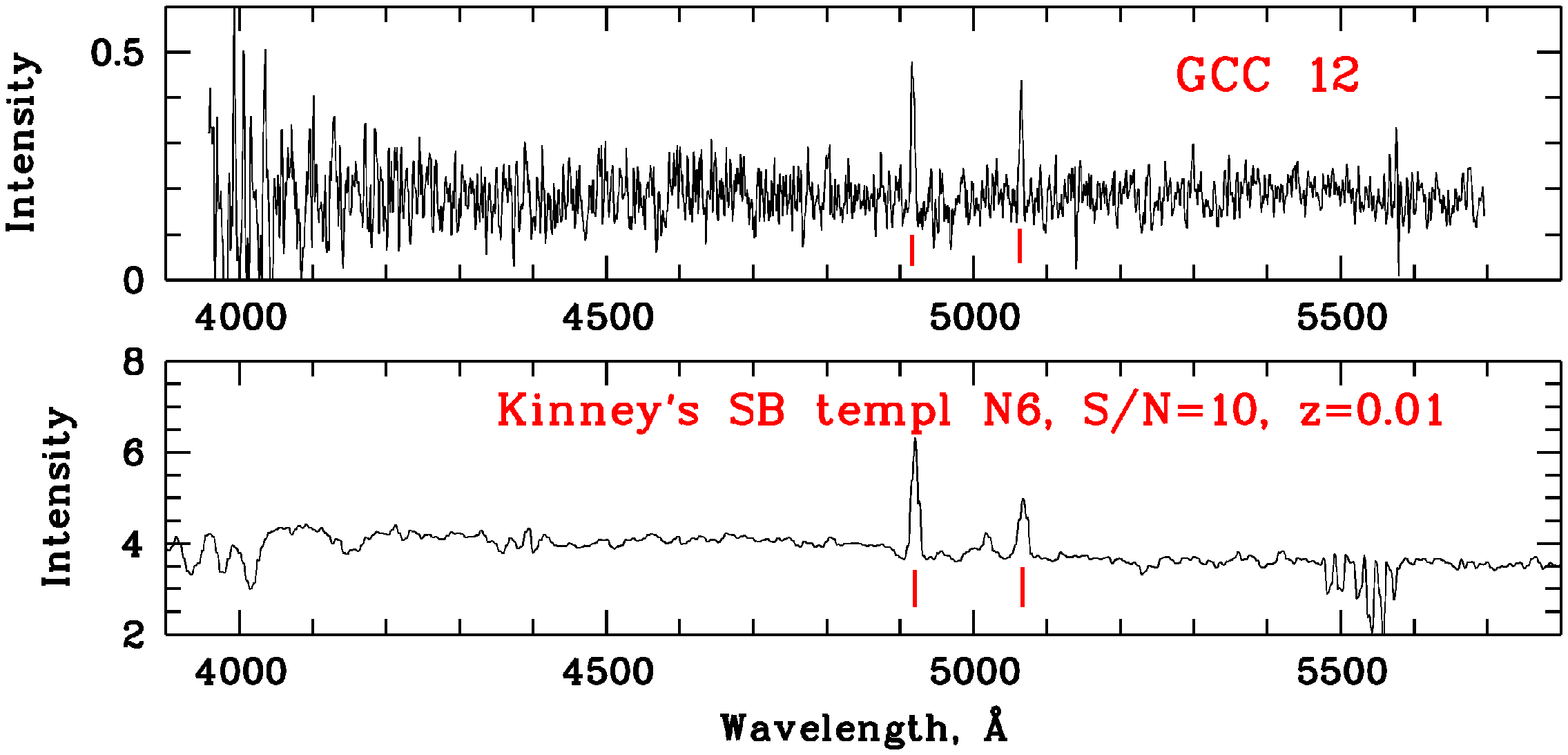}
\caption{Spectra and cross-correlation functions of globular clusters Hodge~IV and GC10, and of GCCs classified as remote galaxies.
 Bright emission-line features in the spectra of distant starburst galaxies ([OII] 3727, $H \beta$, [OIII] 4959, 5007) are indicated.
In these cases templates of Kinney et al. (1996) are shown. They were shifted according to the redshifts.}
\label{ps:corf}
\end{figure*}

\label{ln:obsred}
\begin{table}[!h]
\caption{Images used for GC searches, astrometry, and photometry}
\label{tab:imalog}
\scriptsize
\begin{tabular}{lcllc} \\ \hline\hline\noalign{\smallskip}
Object/Filter  & Date            & Exposure (s)  & Seeing \\ \hline
\noalign{\smallskip}
\noalign{\bf CFHT 3.6m telescope, photographic plates}
\noalign{\smallskip}
NGC~147 (B)           & 16/08/83      & 1800          &  1.0 \\
NGC~147 (V)           & 16/08/83      & 1800          &  1.1 \\
\noalign{\smallskip}
\noalign{\bf Pic-du-Midi 2m telescope, CCD}
\noalign{\smallskip}
Ho III (V, R)         & 02/11/87      & 1800          &  1.4 \\
5$+$6 (B, V, R)       & 16/12/87      & 1200          &  1.0 \\
GC7 (R)                 & 15/12/87      & 720x2         &  1.2 \\
GC7 (R)                 &  16/12/87     &  1800         &  1.2 \\
GCC8 (V)                 & 05/08/88      &1800x2         & 1.1   \\
GCC9 (V)                 & 05/08/88      & 1800          & 1.1   \\
GCC11 (V)                & 07/08/88      & 1800          & 1.3   \\
GCC11 (R)                & 08/08/88      & 1800          & 0.9 \\
GCC12 (V)                & 07/08/88      & 1200          & 1.1   \\
\noalign{\smallskip}
\noalign{\bf 6m telescope of Russian A.S., CCD}
\noalign{\smallskip}
GC10 (R)               &  14/08/07     & 30x2          &  1.2 \\
GC10 (V)               &  03/08/08     & 30x2          &  1.0 \\
GCC12 (R)               & 14/08/07      & 30x2          &  1.2 \\
\noalign{\smallskip}
\hline
\end{tabular}
\end{table}

\begin{table}[!h]
\centering
\caption{Coordinates, magnitudes, heliocentric radial velocities and projected distances 
of GCs and GCCs.
Indices are 1: from Sharina \& Puzia (2009); 2: (V-I) instead of (V-R); 3 : from SAP06.}
\label{tab:coord}
\scriptsize
\begin{tabular}{llllcr} \\ \hline\hline\noalign{\smallskip}
Object & R.A. (2000) Dec.    & V   & $V\!-\!R$  & $V_h,$ & $d_{pr},$ \\
       &                     &     &            & km/s  &  kpc         \\
\hline
\noalign{\smallskip}
GC5 & 00 32 23.4+48 25 44.7 & 18.8& 0.8              & -187 & 2.1        \\
    &                       &     &                  & $\pm$15   &          \\
GC7 & 00 32 22.2+48 31 27.5 & 17.4& 0.8              & -198    & 1.8        \\
    &                       &     &                  & $\pm$10   &              \\
GC10 & 00 32 48.6+48 32 12.4 & 20.5& 0.8              & -180    & 0.9         \\
    &                       &     &                  & $\pm$30     &            \\
H~I & 00 33 12.2+48 30 32.6 & 18.5$^1$  & 1.4$^{1,2}$ & -107$^3$      & 1E-3          \\
    &                       &     &                  & $\pm$30     &            \\
H~II & 00 33 13.6+48 28 48.3 & 18.7$^1$  & 1.2$^{1,2}$ & -189     & 0.7          \\
     &                       &     &                  & $\pm$25                 \\
H~III & 00 33 15.1+48 27 23.4 & 16.8$^1$  & 1.1$^{1,2}$ & -118$^3$ & 0.4           \\
	  &                       &     &                  & $\pm$30                 \\
H~IV  & 00 33 15.0+48 32 09.7 & 19.8& 0.9              & -235    & 0.4           \\
	  &                       &     &                  & $\pm$35                 \\
\noalign{\smallskip}
GCC6 & 00 32 22.9+48 25 48.9 & 17.4& 1.0 & 19200 \\
GCC8 & 00 32 35.3+48 27 28.9 & 19.4 & -  & 27000 \\
GCC9 & 00 32 39.0+48 21 36.0 & 18.5&  -  & 25800 \\
GCC11 & 00 33 04.3+48 26 46.1 & 19.9 & 0.5 & 48000 \\
GCC12 & 00 33 47.2+48 23 24.4 & 19.8&  0.9 & 3000  \\
\noalign{\smallskip}
\hline
\end{tabular}
\end{table}

\subsection{Spectra}
\begin{figure*}[!t]
\includegraphics[width=6.0cm]{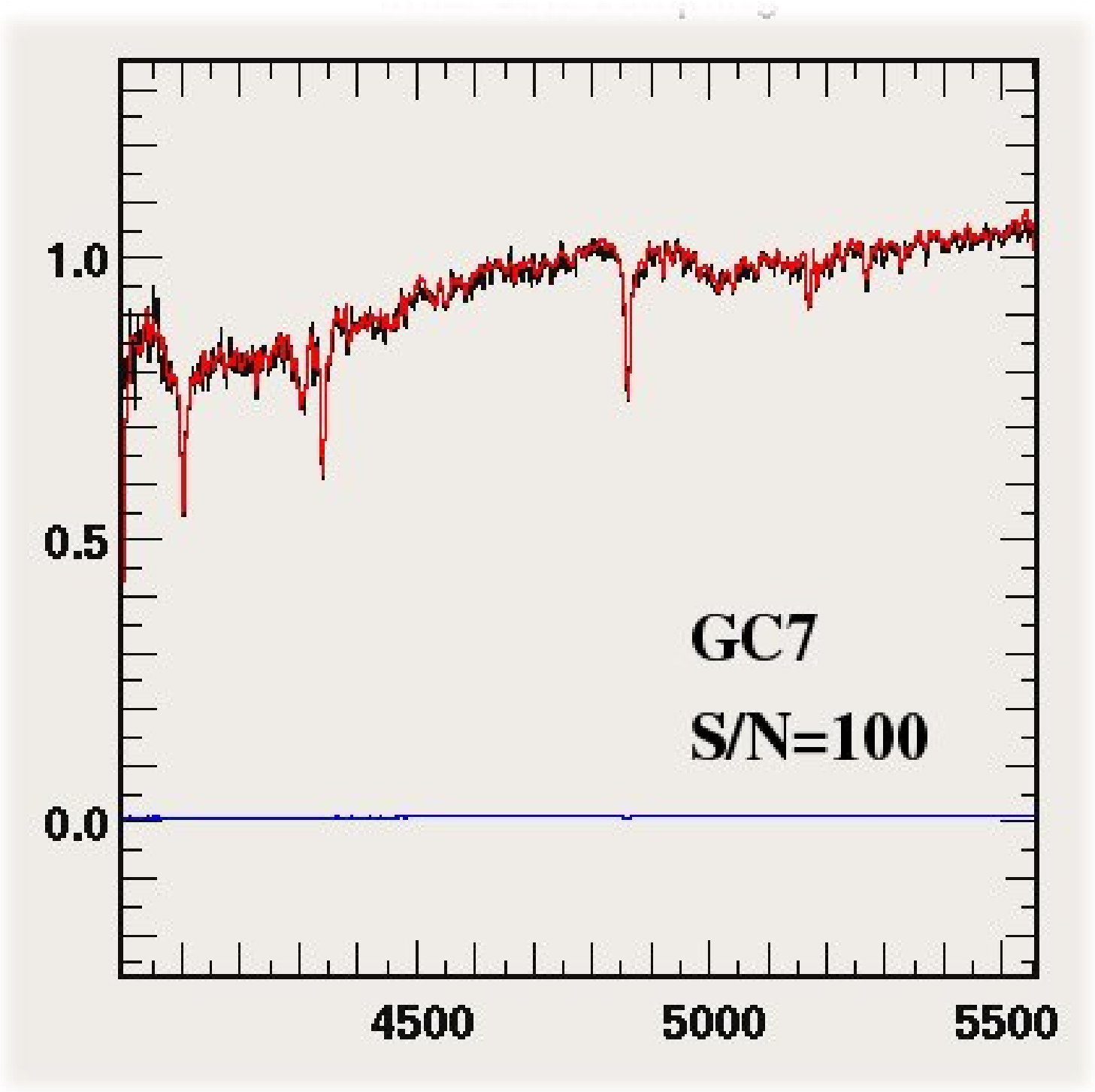}
\includegraphics[width=6.0cm]{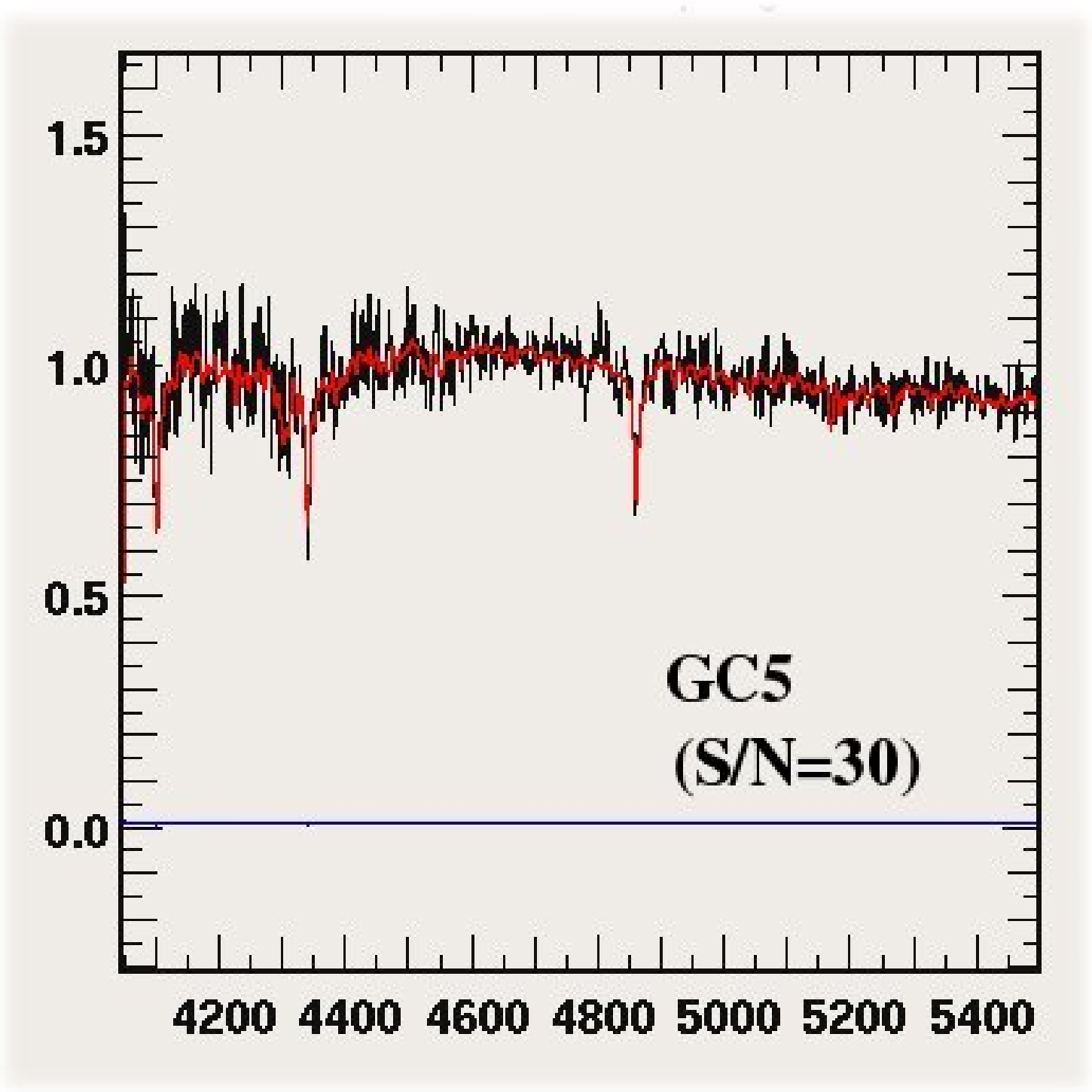}
\includegraphics[width=6.0cm]{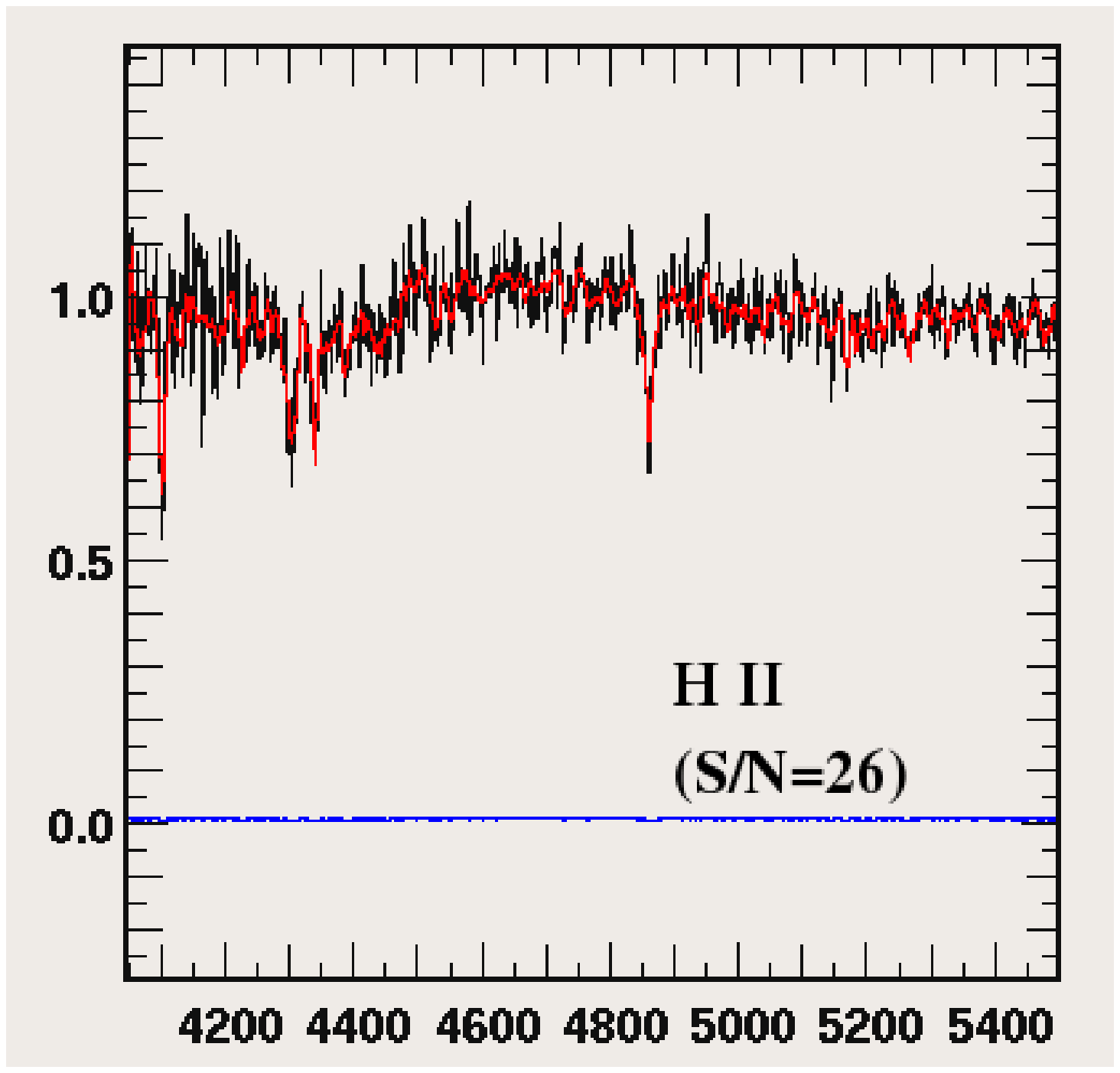}
\caption{ Spectra of GC7, GC5, and Hodge~II scaled to the original continuum at $\lambda=5000$\AA.
PEGASE.HR model spectra fitted by the STECKMAP program (Ocvirk et al. 2006 a,b) are overplotted.}
\label{ps:spectra}
\end{figure*}

\begin{table}[!h]
\centering
\caption{Journal of spectroscopic observations}
\label{tab:obslog}
\scriptsize
\begin{tabular}{lclc} \\ \hline\hline\noalign{\smallskip}
Object         & Date          & Exposure (s)  & Seeing         \\ \hline
\noalign{\smallskip}
GC7          & 11/08/07      & 4x900         &  2.0            \\
Hodge~II$+$GCC11& 11/08/07      & 900           &  2.0            \\
GCC9          & 11/08/07      & 3x900         &  2.0            \\
GC5$+$GCC8  & 13/08/07      &  1200         &  2.0            \\
Hodge~II$+$Hodge~4& 13/08/07    & 2x900         &  1.2            \\
  GC10        & 14/08/07      & 900           &  2.0            \\
  GCC12       & 14/08/07      & 3x900         &  2.0            \\
GCC6          & 03/08/08      & 3x600         & 1.7             \\
GC10          & 03/08/08      & 3x900         & 2.3             \\
\noalign{\smallskip}
HD7010         & 11,13,14/08/07& 20            &  2.0            \\
HD7010         & 03/08/08      & 10            &  2.3            \\
HD2665         & 11,14/08/07   & 10            &  2.0            \\
HD2665         & 03/08/08      & 5             &  2.3            \\
HR1805         & 11/08/07      & 2             &  2.0            \\
\noalign{\smallskip}
GRW+70$^{o}$5824& 11/08/07     & 100           &  2.0            \\
HZ4            & 13/08/07      & 150           &  2.0            \\
\noalign{\smallskip}
\hline
\end{tabular}
\end{table}

The long-slit spectroscopic observations were performed with the SCORPIO
spectrograph (Afanasiev \& Moiseev, 2005), installed at the prime focus of
the 6m telescope of the {\it Russian Academy of Sciences}.
The journal of the spectroscopic observations is given in Table~\ref{tab:obslog}.
We used the CCD detector EEV42-40, the grism VPHG1200g (1200 lines/mm) with a spectral resolution
$\sim\!5$ \AA, the spectral range 3800 -- 5400 \AA. The slit width was 6\arcmin x1\arcsec.
We observed Lick standard stars from the list of Worthey et al. (1994)
for calibration of our instrumental absorption-line system into the Lick
standard one (Worthey, 1994; Worthey \& Ottaviani, 1997).

Although the weather was mostly good during the observing run of 2007,
thin cirrus developed from time to time, and the end of
the night 13/08/07 was lost due to a sudden thunderstorm, preventing us from observing
all the standard stars.
Some exposures were lost in the night 14/08/07 due to disrepair in the filter wheel.
The CCD detector became fogged over in the central part ($\sim 700 \times 700$ pixels) in the final night 2008,
forcing us to use only part of the detector.

The data reduction and analysis were performed using the
European Southern Observatory Munich Image Data Analysis System (MIDAS) (Banse et al., 1983),
and the Image Reduction and Analysis Facility (IRAF) software system\footnote{http://iraf.noao.edu/}.
Cosmic-ray removal was done with the FILTER/COSMIC program in MIDAS.
The standard procedure of primary reduction was applied to each two-dimensional spectrum.
Wavelength calibration was carried out using an arc spectrum taken at the same pointing positions
as the cluster spectrum before or after the object exposure.
The dispersion solution provides the accuracy of the wavelength calibration $ \sim$0.08 \AA.
A typical dispersion was 0.88 \AA/pix. The wavelength zeropoint shifted during the night by up to
2 pixels. It was checked using the [OI]$\lambda5577$ night sky line in the dispersion-corrected spectra.
Extraction of the spectra (Horne, 1986) was made using the IRAF procedure {\it apsum}.

After wavelength calibration and sky subtraction,
the spectra were corrected for extinction and flux-calibrated using
the observed spectrophotometric standard stars GRW+70$^{o}$5824, and HZ4 (Oke, 1990).
Due to non-photometric conditions, the flux calibration is approximate.
Finally, all one-dimensional spectra of each object were summed to increase the S/N ratio.

The radial velocities of the objects were determined using the method of Tonry \& Davis (1979).
We used several radial velocity standards, the PEGASE.HR synthetic spectra\footnote{http://www2.iap.fr/pegase/pegasehr}
for cross-correlation. To estimate radial velocities of remote emission-line galaxies, we used
the template starburst galaxy spectral energy distributions SB6 shifted
to the redshift of the studied objects.
The derived heliocentric radial velocities are listed in Table~\ref{tab:coord}.
The resulting normalized spectra of GCs and GCCs, together with spectra
of standard stars, and cross-correlation functions are shown in Fig.~\ref{ps:corf}.

The flux-calibrated spectra of the two brightest spectroscopically confirmed GCs
and of Hodge~II scaled to their continuum at  $\lambda=5000$~\AA\  are shown in Fig.~\ref{ps:spectra}.

\section{Results}
\label{ln:evol_parm_determ}

\subsection{Methods of age, metallicity and $\alpha$-element abundance ratio determination}
\begin{figure*}[!t]
\centering
\includegraphics[width=12.5cm]{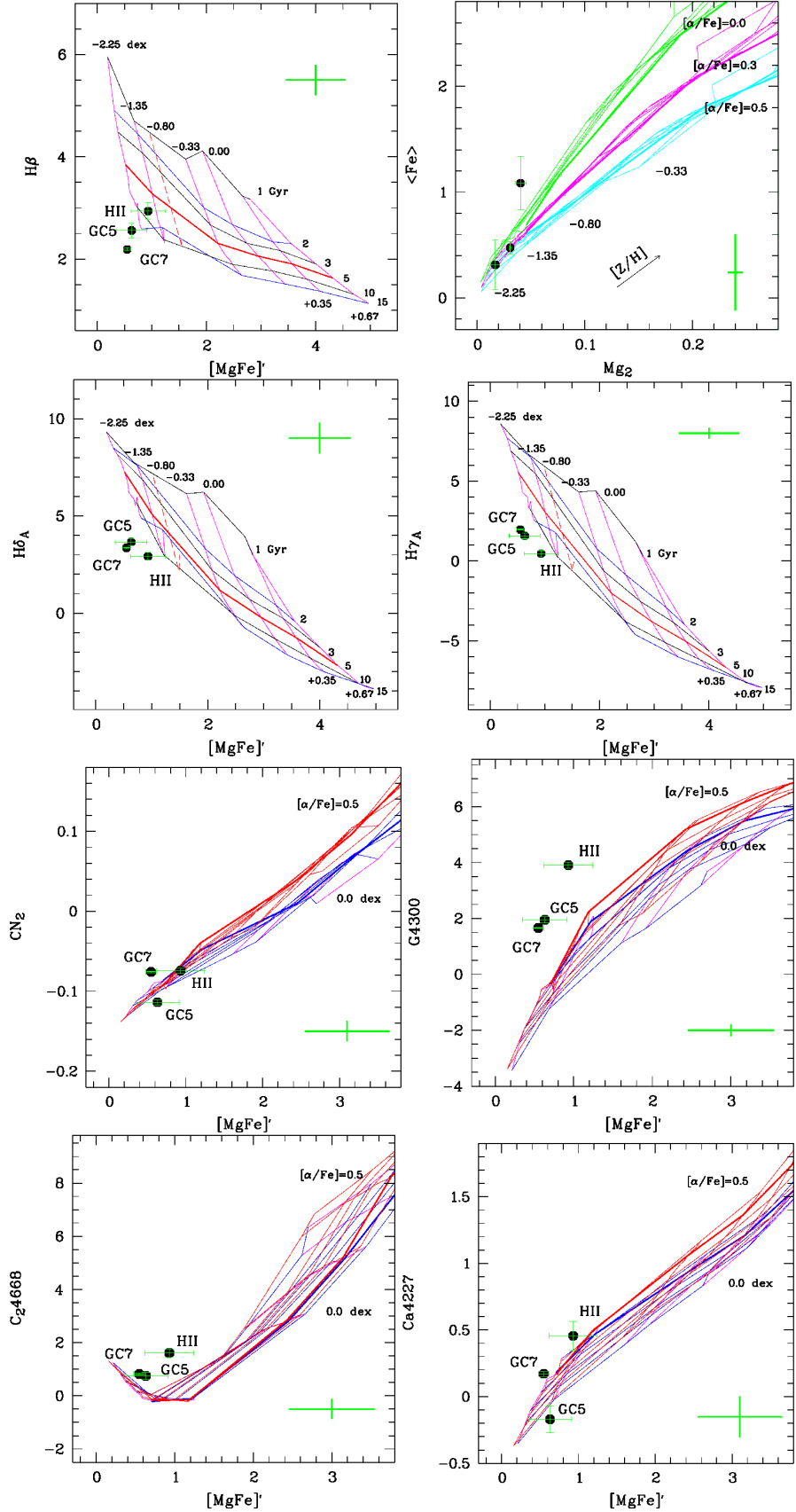}
\caption{Diagnostic plots of age-metallicity, \afe - metallicity, and [MgFe]\arcmin versus indices sensitive to C, N, and Ca. The bootstrap errors are
overplotted. We use the SSP model predictions from Thomas et al. (2003, 2004).
Error bars on each plot show rms errors of the transformations to the Lick system.}
\label{ps:diag_plots}
\end{figure*}

In the analysis of the medium-resolution spectra, 
we applied three methods for determining the evolutionary parameters.
The first one is our main working method (see also SAP06) for comparing
intensities of absorption-line features in spectra of GCs using the Lick-index definitions of
Worthey (1994) and Worthey \& Ottaviani (1997) with predictions of SSP models (Thomas et al. 2003, 2004).
 Two additional approaches proposing full spectrum fitting using the PEGASE.HR model spectra (Le Borgne et al. 2004) helped us understand how
the choice of models, the number of spectral features involved in the analysis,
imperfection of the flux calibration, and uncertainties of the Lick index zeropoints influence
the derived parameters. 
 Anticipating on the results, we point out that the three methods mentioned above gave
similar ages and metallicities within the errors for
the GCs\footnote{We use the standard definition, $[X/Y]=log(X/Y)- log(X_{\sun}/Y_{\sun})$,
where X and Y are masses of specific elements. \zh\ is the overall metallicity.}.
In the following we describe the methods and results in detail.
\begin{table}[!h]
\centering
\caption{Correction terms of the transformation to the Lick/IDS
standard system: $ I_{\rm Lick}=I_{\rm measured}+c $.}
\label{tab:zeropoints}
\begin{tabular}{lrccr} \\ \hline
Index       & c          & rms error & Index range& units \\ \hline
\noalign{\smallskip}
CN1         & $-$0.060   & 0.033 &  [-0.07 -- 0.4]   & mag    \\
CN2         & $-$0.016   & 0.013 &  [-0.05 --0.4]    & mag    \\
Ca4227      & $-$0.06    & 0.154&   [0.17 -- 2.5]    &  \AA   \\
G4300       & 0.23       & 0.219&   [5.2 -- 7.0]     &  \AA   \\
Fe4384      & $-$0.08    & 0.267&   [2.3 -- 4.0]     &  \AA   \\
Ca4455      & 0.65       & 0.299&   [1.0 -- 9.1]     &  \AA   \\
Fe4531      & $-$0.07    & 0.609&   [1.0 -- 5.2]     &  \AA   \\
Fe4668      &    0.20    & 0.773&   [0.2 -- 10.4]    &  \AA   \\
H$\beta$    &$-$0.30     & 0.472&   [0.8 -- 1.1]     &  \AA   \\
Fe5015      & 0.30       & 0.755&   [1.6 -- 7.3]     &  \AA   \\
Mg$_1$      & $-$0.002   & 0.029&   [-0.01 -- 0.21]   &  mag    \\
Mg$_2$      & 0.020      & 0.077&   [0.01 -- 0.36]   &  mag    \\
Mgb         & 0.10       & 0.400&   [0.3 -- 3.9]     & \AA  \\
Fe5270      & 0.22       & 0.746&   [0.7 -- 4.3]     & \AA  \\
Fe5335      & 0.40       & 0.700&   [0.1 -- 3.9]     & \AA  \\
Fe5406      &  0.20      & 0.518&   [0.2 -- 2.9]     & \AA  \\
H$\delta_A$ & $-$0.74    & 0.841&   [-7.6 -- 0.6]     & \AA  \\
H$\gamma_A$ & $-$0.06    & 0.026&   [-11.6 -- -3.6]     & \AA  \\
H$\delta_F$ &  $-$0.30   & 0.076&   [-2.1 -- 1.1]     &  \AA  \\
H$\gamma_F$ &  $-$0.07   & 0.154&   [-3.5 -- -0.7]     &  \AA  \\
\noalign{\smallskip}
\hline
\end{tabular}
\end{table}

\subsubsection{Lick indices}
\label{ln:lick_indices}
\begin{table*}[!hbt]
\caption{Ages, metallicities and \afe\  for GCs with high S/N ratio obtained using three different methods
(see Section~3.1 for details).}
\label{tab:evolpar}
\begin{center}
\begin{tabular}{lccc|cc|cr} \\ \hline\hline\noalign{\smallskip}
Object & Age$_{{\chi}^2}$& \zh$_{{\chi}^2}$ & \afe$_{{\chi}^2}$& Age$_{STECK}$   & \feh$_{STECK}$& Age$_{CS}$      & \feh$_{CS}$      \\
       &  Gyr            &  dex             &   dex            & Gyr          &  dex           & Gyr          &  dex          \\
\hline\noalign{\smallskip}
GC7   & 8$\pm$2         & -1.5$\pm$0.2    & 0.2$\pm$0.2      & 8            &  -1.3      & 6$\pm$2    & -1.6$\pm$0.2 \\
GC5   & 10$\pm$2        & -1.7$\pm$0.2    & 0.1$\pm$0.2      & 10           &  -1.5      & 10$\pm$3   & -1.7$\pm$0.3 \\
Hodge~II& 9$\pm$3        & -1.1$\pm$0.3    & 0.5$\pm$0.4      & 7            &  -1.2      & 10$\pm$4   & -1.2$\pm$0.4 \\
\noalign{\smallskip}
\hline
\end{tabular}
\end{center}
\end{table*}

The most commonly used method for disentangling age-metallicity degeneracy effects
is that of the Lick indices (e.g. Worthey et al., 1994; Worthey, 1994; Worthey \& Ottaviani, 1997;
 Faber, 1973; Proctor et al., 2004; Puzia et al., 2005b).
The quality of the spectra allowed us to accurately measure the absorption-line 
indices and estimate ages, metallicities, and $\alpha$-element abundance ratios ([$\alpha/Fe$])
of our GCs by comparison of the measured values with the model ones.

We used a number of Lick standard stars observed with the
long-slit unit at the SCORPIO spectrograph from previous programs to
calculate zeropoints of transformation of our Lick index
measurements into the standard system (Worthey, 1994) (see Table~\ref{tab:zeropoints}).
The range of index values covered by our standard stars is given in column~4 of  Table~\ref{tab:zeropoints}.

The calculated
zeropoint transformations to the Lick/IDS system agree with those given in Sharina et al. (2006a).
However, we point out that we managed to obtain spectra for only 3 Lick standard stars during the
observing runs of 2007, 2008.
The rms errors of zeropoints are consequently large. So, for better confidence in the results
we used here additional methods of evolutionary parameter determination,
independent of the Lick index one, (see subsections~\ref{ln:steckmap}, ~\ref{ln:chi} for details).

Absorption-line indices for the spectroscopically confirmed GCs with S/N$\ge 25$,
measured by the GONZO program (Puzia et al., 2002) and transformed into the Lick system
are listed in Tables~C.1, C.2.  The spectra were degraded to the
resolution of the Lick system before measuring the indices.

To estimate the evolutionary parameters we use a three-dimensional interpolation and $\chi^2$
approximation routine (SAP06), and SSP models of Thomas et al (2003, 2004).
 The $\chi^2$ minimization technique has the advantage of using
all available indices to determine the evolutionary parameters (Proctor et al., 2004).
This reduces the influence of individual index uncertainties on the results.
Systematic errors arise because of variable atmospheric transparency, instrumental problems,
and imperfections of data reduction:
uncertainties of flux calibration and sky subtraction, scattered light effects,
spectral wavelength calibration and radial velocity determination,
corrections from spectral resolution, seeing, focus and velocity dispersion,
response deviations from linearity, the contribution from nebular absorption lines.
Uncertainties of transformation into the Lick system appear to be significant
when the number of observed standard stars is small.
Statistical index errors are due to random Poisson noise.
A typical error of metallicity determination using this approach for spectra of
GCs with S/N$\ge$30 is less than 0.2 dex  (SAP06).
Relative errors of age determination are  $\sigma_t = \delta(log(age/yr)) \sim 0.2$.
The results of evolutionary  parameter determination using this method
are presented in columns 2 -- 4 of Table~\ref{tab:evolpar},  and are illustrated in Appendix~B
as $\chi^2$ and confidence contours in the age -- \zh, \afe -- \zh, and age -- \zh\ planes.
A description of the program is also given.

We derive low metallicities ($\feh \le -1.1$ dex), and old ages (8-10 Gyr) for GC5, GC7, and Hodge~II
in good agreement with the evolutionary parameters for other GCs in NGC~147 (Hodge~I, III) reported by SAP06.
A spectrum of Hodge~II of low signal-to-noise ratio (S/N$=14$) was obtained by SAP06\footnote{Note that
the identifiers of Hodge~II and III were inverted by mistake in SAP06.}.
They obtained different values for the metallicity (\zh=-1.5 dex), and age (10 Gyr).
The $\alpha$-element ratio for Hodge~II is not very accurate in this paper and in SAP06.
It has a high value of 0.5 dex with an uncertainty of 0.4 dex in both cases.

If one compares the location of the measured Lick indices with model predictions
on two-dimensional index plots, it appears that some of the line strengths are not
well reproduced by the models. Figure~\ref{ps:diag_plots} shows age -- metallicity,
\afe -- metallicity, and carbon, nitrogen, and calcium sensitive indices versus metallicity
diagnostic plots (Puzia et al., 2005b).
Metallicity indices insensitive to \afe\ are used to construct them, namely 
[MgFe]\arcmin~$= \left\{ \mbox{Mg}b \cdot (0.72 \cdot \mbox{Fe5270} + 0.28 \cdot \mbox{Fe5335})\right\}^{1/2}$,
and $\langle Fe \rangle = (Fe5270 + Fe5335)/2$. The Balmer line indices (H$\beta$, H$\gamma_A$, H$\delta_A$)
are sensitive to the temperature of
hot main-sequence turn-off stars, metal lines arise from the coolest red-giant branch and lower 
main-sequence stars, and the Mg$_2$ index is very sensitive to the $\alpha$-element abundance ratio
(Worthey, 1994; Worthey \& Ottaviani, 1997).
This is why the H$\beta$--[MgFe]\arcmin, H$\gamma$--[MgFe]\arcmin, H$\beta$--[MgFe]\arcmin,
and $<Fe>$--Mg$_2$ diagnostic plots allow one to define the evolutionary parameters (Puzia et al., 2050b).

CN$_2$ is sensitive to carbon abundance, while the G4300, C$_2$4668 line strengths are also
sensitive to C, but not sensitive to nitrogen variations (Tripicco \& Bell, 1985).
The GCs show insignificant deviations from the model
predictions in the CN$_2$--[MgFe]\arcmin\ diagram. However, they show high G4300, C$_2$4668 line strengths.
This may indicate that the GCs are overabundant in C and underabundant in N.
This qualitative result needs confirmation with higher resolution spectroscopic observations.

Figure~\ref{ps:diag_plots} shows that the
Balmer-line indices fall below the oldest model sequences,
which may indicate ages older than the age of the Universe.
For some GCs, the age derived from one diagnostic diagram differs from
that derived from the other diagrams.
The reason lies not only in random and systematical errors in measuring
Lick indices, but also in probable SSP model problems.

The latter are ultimately tied to different physical conditions
in dwarf galaxies, where we study GC systems, and in our Galaxy,
where stars are used to construct the grids.
Giant and dwarf galaxies experience very different evolutionary histories,
which influence their stellar mass functions and chemical composition.
Stars in dwarf galaxies are metal poorer, show lower $\alpha$-element abundance
ratios, higher s-process element to Fe ratios (Venn et al. 2004, Pritzl et al. 2005).
Supernovae of type Ia start to contribute at lower metallicities and at earlier ages
in dwarfs in comparison to their massive neighbours.
Dwarf galaxies experience a slower chemical evolution with
possible short periods of higher star formation rates at young ages.
As a result, specific element abundances may be observed (Venn \& Hill, 2008).
Additionally, it is worth noting that the stellar atmospheres interact with the
interstellar medium, whose low temperature in dwarf galaxies may influence
the production of resonance lines (CaI, NaI) with very low ionization potentials
by definition, and molecules (CN, CH, C$_2$), on which the corresponding Lick indices,
Ca4227, NaD, and CN$_1$, CN$_2$, G4300, C$_2$4668, are centered.
These absorption features are weakened much more than others (e.g. MgI, FeI)
with the temperature growth.


\subsubsection{STECKMAP results}
\label{ln:steckmap}

We used STECKMAP (STEllar Content and Kinematics via Maximum a Posteriori)
(Ocvirk et al. 2006 a, b) as an independent method for determining evolutionary parameters.
This non-parametric method is intended to recover kinematic properties and the stellar content
of galaxies from their integrated-light spectra.
It has also been applied to Galactic GCs (Koleva et al., 2008).
It allows one to determine ages and metallicities of stellar
systems from the corresponding parameter distributions
by summing luminosity-weighted age and \feh\  fractions for each SSP model bin:
$$Age = \sum_{i=1}^{n} X_i t_i; \hspace{5mm} \feh = 10 \sum_{i=1}^{n} X_i log(\feh_i), $$
where $t_i$ and $\feh_i$ are the age and the metallicity of the i-th SSP of the basis,
and $X_i$ is the weight (i.e. flux fraction) of this component.

The results are presented in Table~\ref{tab:evolpar} (columns 5, 6). Normalized spectra of three brightest GCs fitted
using the STECKMAP program are shown in Fig.~\ref{ps:spectra}.

\subsubsection{ Continuum shape fitting}
\label{ln:chi}

We were not aiming at an absolute flux calibration in our spectra, because the methods of analysis we use
do not require it. A critical condition for using the above two methods is a high
signal-to-noise ratio in the spectra (S/N $\ge 30$). 
Atmospheric transparency was slightly variable during our observations.
So we decided to make use of a possible imperfection in the absolute flux calibration
as an additional way of determining the evolutionary parameters.

\begin{figure}[!t]
\centering
\includegraphics[width=6cm, angle=-90]{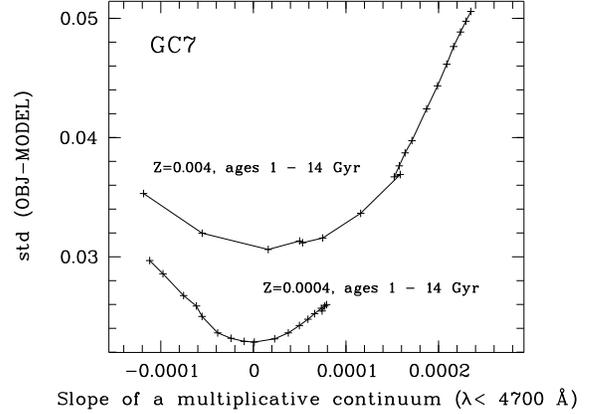}
\caption{Result of division of the blue continuum slopes ($\lambda < 4700$ \AA) in the spectra of GC7
and two sets of the PEGASE-HR models of metallicities Z=0.004, and Z=0.0004 and different ages (1 -- 14 Gyr)
versus the difference between the object and model spectra, both normalized and scaled to the flux at $\lambda < 5000$ \AA.}
\label{ps:chi2}
\end{figure}

In the following we compare our spectra of GCs with those of the PEGASE.HR models
and check whether the continuum slopes and intensities of
different absorption lines in our spectra and the model ones are correlated,
and how they are influenced by age and metallicity.
Note that, as when using the STECKMAP program, we take into account the whole spectra
with all the information they contain.
Our task is to find two quantities for each observed spectrum and each model age and metallicity:
the slope of a multiplicative term to be applied to a GC spectrum to bring it
into correspondence with a model one, and the difference between the model and the transformed object spectra.
For this we first scale model and object spectra at $ \lambda = 5000$~\AA\ and normalize simultaneously.
The ratio of the normalization terms defines the miltiplicative continuum:
$ Z_N = \frac{modelCONT_N}{objectCONT_N}$, $ OBJ_N = obsOBJ_N \cdot Z_N$,  where $N$ is a spectral element.
Then we find the difference between the object spectra scaled at $ \lambda = 5000$ \AA\ and multiplied by $ Z_N$
and the corresponding model ones.
Figure~\ref{ps:chi2} shows the dependence between two parameters, the slope of $Z_N$ in its blue
part ($\lambda < 4700$ \AA\ ) and  $\sigma(OBJ_N - MODEL_N)$ for two sets of the PEGASE-HR model spectra:
1)~Z=0.004, ages from 1 to 14 Gyr, and 2)~Z=0.0004, ages from 1 to 14 Gyr.
The minimum standard deviation of the difference between the resulting object and model spectra at the minimum
slope indicates the age and metallicity characterizing our object.

Linear interpolation between the SSP model grids and
consideration of  $ \chi^2 = \sum_{N} \frac{(OBJ_N - MODEL_N)^2}{\sigma(OBJ_N)^2} $ parameter distribution with
the observed spectrum of errors, $ \sigma(OBJ_N) $, allow one to estimate errors of age and metallicity determination.
This approach is a simplification of methods used for modeling the spectral evolution of galaxies
(Chilingarian, 2007; Koleva et al., 2008).

 The results of our spectrum fitting are presented in Table~\ref{tab:evolpar} (columns 7, 8).

\subsection{Specific frequency of GCs in NGC~147}

The number of globular clusters per unit $M_V\!=\!-15$ mag of host galaxy light, $S_{N}\!=\!N_{\rm GC} \cdot 10^{0.4(M_V+15)}$
(Harris \& van den Bergh, 1981), is equal to 6.4 for NGC~147.
This value is higher than the one expected if a constant number of GCs per unit mass
are formed with a constant efficiency: $S_{N} \sim 0.0025\, L_{\rm gal}^{0.3}$ (Mclaughlin, 1999).

The corresponding number of GCs per unit galaxy mass $M_G=10^9 M_{\sun}$ is $$ T = \frac{N_{GC}(tot)}{M_G/10^9 M_{\sun}},$$
(Zepf \& Ashman, 1993).
Using the mass estimate $5 \cdot 10^8 M_{\sun}$ (see Appendix~A), $T = 14$ for NGC~147.
This value is higher than for NGC~205 ($T=4$), and NGC~185 ($T=12.5$) using the mass estimates of Bender et al. (1991).
However, it is much lower than for dark matter dominated dSphs. For example, 
the extremely low surface brightness dSph KK221 in the Centaurus~A group 
has a mass $M \sim 6 \cdot 10^7 M_{\sun}$
and a number of spectroscopically confirmed GCs $N_{GC}=6$ (Sharina \& Puzia, 2008), 
thus the mass fraction of GCs in that galaxy is $T = 100$.

\section{Discussion}

In order to patch together a reasonable scenario for the formation history of NGC~147,
we now compare the physical parameters of its GCs with those of the field stars, 
as well as with those of stars and GCs in other galaxies of the M31 group.
\subsection{NGC~147}
\subsubsection{Metallicities and abundance ratios}
The evolutionary status of diffuse stellar populations in NGC~147 
has been studied by many authors starting with
Mould, Kristian, \& Da Costa (1983), who estimated a mean metallicity of 
\zh$=-1.2 \pm 0.2$ dex, with a spread of 0.3 dex, for the red giants in the outer parts of the galaxy.
This coincides with the value in
the stellar halo of M31 at galactocentric distances greater than 60 kpc (Kalirai et al., 2006).
In the center of NGC~147, the metallicity is higher : \feh=-0.9 dex (Grebel, 2000; McConnachie et al., 2005; Dolphin, 2005).
The mean abundance of planetary nebulae, $\feh \sim -0.97$ (Gon\c{c}alves et al., 2007), is close to the one of the old stellar populations.

By comparison, the bright and massive GCs of NGC~147 have a very low mean metal content, 
much lower than the halo stars
(Da Costa \& Mould, 1988; SAP06; Table~\ref{tab:evolpar} in this paper).
The faint GCs Hodge~I and Hodge~II near the center of the galaxy show enhanced metal abundances, 
close to the metallicity of the oldest stars ($\zh \sim -1.0 \div -1.2$ dex).
 All the GCs have solar $\alpha$-element abundance ratios within the errors.

\subsubsection{Ages}
NGC~147 contains a small amount of intermediate-age C and M stars 
(Nowotny et al., 2003; Davidge, 1994, 2005).
The most recent star-forming event in NGC~147 happened $\sim 3$ Gyr ago, according to 
broad-band near-infrared color-magnitude diagrams (Davidge, 2005).
On the other hand, the absence of main-sequence turn-off stars 
with $M_V < -1$ indicates that the most recent large-scale
star-forming activity occurred at least 1 Gyr ago (Han et al., 1997).
It may have coincided with the formation of the brightest M giants (Sohn et al., 2006).
These estimates of the most recent star-forming event agree with the study
of Dolphin (2005) using HST images.
It is important that the brightest AGB stars in NGC~147
are well-mixed with fainter stars. This may indicate that the whole galaxy was involved in the
star formation process, or that it has been tidally stirred by interactions with another galaxy.

The GCs in NGC~147 appear to be much older than the brightest AGB stars (see Table~7 in SAP06 and Table~\ref{tab:evolpar} this paper).
Their mean age is 9$\pm$1 Gyr.
It should be noticed, however, that the brightest clusters GC7 and Hodge~III,
despite their low metallicities, show rather young ages, 
similar to those of the more metal-rich clusters Hodge~I, and II.
This may be the result of artificial filling-in of the Balmer lines, caused by hot stars, 
such as horizontal branch (HB) and emission-line stars
(Schiavon et al., 2005; Recio-Blanco et al., 2006). These artificially younger ages may also result from the
presence of multiple stellar populations, since Balmer lines are mostly sensitive to the main-sequence
turn-off temperature (Worthey, 1994). In our Galaxy, the most massive GCs also show indications of unusually hot,
extended HBs, multiple stellar populations, and are kinematically distinct from other Galactic
GCs (Recio-Blanco et al., 2006; Lee et al. 2007).

\subsection{M31 and other dE satellites}

The absence of young or even intermediate-age GCs in NGC~147 contrasts with the situation in other
dEs of the M31 group.
NGC~185 and 205 contain GCs of intermediate age and metallicity
(SAP06): FJJVII (\zh=-0.8, age=7 Gyr) in NGC~185,
Hubble~V (\zh=-0.6, age=1 Gyr), and Hubble~VI (\zh=-0.8, age=4 Gyr) in NGC~205.
Such younger objects also exist in the M31 halo (see Sec~4.2 in SAP06 and references therein).
Fan et al. (2006) found young and intermediate-age GCs in M31, peaking at $\sim3$ and 8 Gyr
from multicolor photometry.
Spectroscopic studies have shown that three types of GC populations exist in M31: old objects with
a wide range of metallicities, intermediate age and metallicity GCs, and young GCs with slightly higher
metallicity (Perrett et al., 2002; Beasley et al., 2004, 2005; Fusi Pecci et al., 2005; Puzia et al., 2005a).

Stellar populations of intermediate ages and metallicities are also present in the close neighbourhood of M31.
Mean metallicities of stars and metallicity spreads are greater in NGC~205, and 185, respectively
$-0.5\pm0.5$, and $-0.8\pm0.4$ (Grebel, 2000).
The spectroscopic age of M32 is 2 -- 3.5 Gyr (Schiavon et al., 2004).
Brown et al.(2003) estimated that $\sim$30\% of the halo old population in M31 comprises intermediate-age
stars with ages of 6--8 Gyr, and metallicities of $\sim$-0.5 dex.

The spectroscopically studied M31 GCs are poorer in $\alpha$-elements than Galactic GCs: $\afe=0.14 \pm 0.04$ dex
(Pritzl et al., 2005; Puzia et al. 2005a), and
GCs in the M31 group dEs have near-solar $\alpha$--element abundance ratios (SAP06).

\subsection{How did NGC~147 evolve?}

All the aforementioned literature data about ages and metallicities of stellar populations indicate
that there were two main star forming events in the neighbourhood of M31:
$\sim\!1\div 3$ and $\sim\!8$ Gyr ago. In spite of the large errors associated with the evolutionary parameters,
the ages of GCs in NGC~147, 185 and 205 concentrate around these two periods.
Detailed theoretical and observational studies will help to identify the cause of this bimodality.
It might be the tidal disruption of a satellite galaxy, or interactions between  M31 and M33.

The globular cluster system of NGC~147 formed during early and very
powerful star forming events in the life of the galaxy (Da Costa \& Mould, 1988).
From the age spread of the GCs, this active period started $\sim 8 \div 10$ Gyr ago, and continued for $\sim$2 Gyr.
The massive clusters Hodge~III, GC5 and GC7 have lower metallicities than the mean metallicity of metal-poor
stars in the outer parts of NGC~147 and in the M31 halo.
This is an indication of very early formation. 
The faint clusters Hodge~I and Hodge~II and the majority of the halo stars
formed later and simultaneously, since their metallicities are similar.
The difference in metallicities between the youngest low-mass
GCs and the youngest stars is only $\sim 0.3$ dex.
It is thus likely that the gas content was already low after the first starburst, and 
the metallicity has not changed much since.  Finally, after a weaker period of star formation
approximately 3 Gyr ago (Davidge, 2005), the galaxy was depleted of the rest of its gas.

It is interesting that all the outer GCs are located west and south-west of the galaxy center.
This may result from tidal interaction or ram-pressure stripping due to motion through dense gas,
 or else from anisotropic galactic-scale winds, which would also explain the gas depletion in
the galaxy.

\section{Conclusion}

We have reported on the discovery and spectroscopic study of three new GCs in NGC~147.
Medium-resolution spectra were obtained at the 6-m telescope, and radial velocities
were measured for all the objects. The clusters GC5, GC7, and GC10  have heliocentric radial velocities
close to that of NGC~147, while other candidates appear to be remote galaxies.
The newly found GCs are anisotropically distributed: they are located
at projected distances 0.9 -- 2.1 kpc south-west of the center of the galaxy.

We obtained evolutionary parameters for the brightest
globular clusters GC5 and GC7, and for Hodge~II. 
They appear to be old and metal-poor.
Although the evolutionary parameters, derived using medium-resolution spectra, have large uncertainties,
the ages and metallicities of GCs in NGC~147 show clear correlations
with those of red giant branch stars in the galaxy and of GCs and old stars in the M31 halo.
Thus the bright clusters Hodge~III, GC5, and GC7 were likely to be formed first. Their metallicities are
lower than the mean metallicity of the old stars in the galaxy: \zh$\sim -1.5 \div -1.8$.
The fainter clusters Hodge~I and Hodge~II are second generation objects. They show
metallicities similar to that of the oldest stars in the galaxy, \zh$=-1.2$.
We suggest that the main star forming period occurred $\sim$8-10 Gyr ago.
 Then the activity was slow, and the rest of the gas was depleted $\sim1 \div 3$ Gyr ago.

Hodge~III, and GC7, the brightest GCs in NGC~147, seem rather young, given
their low metallicities. We explain this by the possible presence
of multiple stellar populations and unusually hot HB stars.

The mass of NGC~147 from the projected positions and radial
velocities of the GCs is $\sim 5 \cdot 10^8 M_{\sun}$, 
in good agreement with previous estimates. The updated specific frequency and mass
fraction of GCs in NGC~147 are higher than those of NGC~185 and NGC~205.

Deep color-magnitude diagrams and high-resolution spectra of GCs obtained with telescopes in space
will be needed to further progress in our understanding of the evolutionary history of NGC~147.

\acknowledgements
Based on observations obtained at the Canada-France-Hawaii Telescope
(CFHT) which is operated by the National Research Council of Canada, the
Institut National des Sciences de l'Univers of the Centre National de la
Recherche Scientifique of France, and the University of Hawaii, at the
Telescope Bernard Lyot of Observatoire du Pic du Midi,
operated by INSU (CNRS), and at the 6-meter telescope of the Russian
Academy of Sciences.
MES gratefully acknowledges financial support from the Centre National de la Recherche Scientifique (France),
Russian Foundation for Basic Research grants 08-02-00627, and 08-02-08-536$_3$, and deeply thanks Observatoire Midi-Pyr\'en\'ees for its hospitality.
We thank G\'erard Leli\`evre for help with the CFHT observations, and J.F. Le Campion for determining accurate positions for our GCCs;
Dmitrij Makarov, Gretchen Harris, Thomas Puzia, Slava Shimansky and an anonymous referee for valuable comments;
Pierre Ocvirk and Thomas Puzia for help in using of their programs STECKMAP and GONZO.
Preliminary photometric and structural investigations of the data were done by Thierry Devili\`ere.
We are grateful to the Large Telescopes Program Committee of the Russian Academy of Sciences
for allocation of observing time.

{}

\clearpage
\appendix
\section{Mass estimate for NGC~147}
In the following we estimate the mass of NGC~147 from the projected position and line-of-sight velocity of GCs obtained in this
paper and by SAP06. We use the "tracer mass estimator" (Evans, 2003),
applied when the tracer population does not follow the dark matter density profile.
In the case of an isothermal potential:
\begin{equation}
M_{\rm press}=\frac{C}{GN}\sum_i (v_{i, {\rm los}}-\langle v\rangle)^2 R_i,
\end{equation}
where
\begin{equation}
C=\frac{16(\gamma-2\beta)}{\pi
(4-3\beta)}\cdot\frac{4-\gamma}{3-\gamma} \cdot\frac{1-(r_{\rm
in}/r_{\rm out})^{3-\gamma}}{1-(r_{\rm in}/r_{\rm out})^{4-\gamma}}.
\end{equation}
Here, $\langle v\rangle$ is the system's mean radial velocity and $\beta$ the anisotropy parameter
$1-\sigma_t^2/\sigma_r^2$, equal unity for purely radial orbits
and $-\infty$ for a system with solely tangential orbits (Binney, 1981).

We used the power-law rule $\gamma = 1+d\log\Sigma / d\log R$ (Gebhardt, et al. 1996) to derive the three-dimensional
density profile of the GC population. If the surface-brightness profile  is purely exponential, then $\gamma = 2$ (De Rijcke et al., 2006).
We estimate the Sersic indices n along the major and minor axes the galaxy to be $n=0.86$, and $n=1.0$, respectively, using the surface-brightness profiles  (Kent, 1987).
These values are in a good agreement with the conclusion of De Rijcke et al. (2006).

 There are different estimates of the rotation velocity for NGC~147 in the literature. According to De Rijcke et al. (2006),
it is equal to zero. Bender et al. (1991) found rotation with an amplitude $V_r\sim10$ km/s.

The mass of NGC~147 is equal to $M_{\rm press} \approx 6.2\pm2 \cdot 10^8 M_{\sun}$ from equations (A.1) and (A.2)
in the case of $V_r\sim0$ km/s, the anisotropy parameter $\beta=0.5$ for randomly oriented orbits, and radii of orbits
of the nearest and most distant GC: $r_{\rm in}=1.2$ pc and $r_{\rm out}=2.08$ kpc.
The statistically unbiased estimate of the mass value is
$M_{\rm press}^c= M_{\rm press}[1-(2 \sigma_v^2)/3 s_v^2] \sim M_{\rm press} \cdot 0.8 \sim 5 \cdot 10^8 M_{\sun}$ ,
where $\sigma_v$ is the rms error of the radial velocity measurements, and $s_v$ is the
rms velocity of GCs relative to the mean velocity of the GC system, $s_v^2= (1/k)\sum{(v_k-\langle v \rangle)^2}$
with a number of GCs equal k (Karachentsev et al., 1999).

The expected uncertainty of the total mass estimate is $\sim\!50$\% taking into account the small number of GCs.
In the present case of seven GCs, the error on the velocity
dispersion measurement is $\sim\!23$\% of the value of the velocity dispersion itself.
In  spite of this large uncertainty the obtained mass and $M/L_B$ estimate are in good
agreement with those of Bender et al. (1991).

\section{Comparison of our Lick index measurements with SSP model
predictions using a 3-dimensional interpolation and $\chi^2$
minimization routine}

 In this section we illustrate the results described in Section~3.1.1.

With our method (see also SAP06), we first obtain a full set of theoretical Lick indices for any age, metallicity
and \afe\ via three-dimensional linear interpolation. Then we minimize the following $\chi^2$ function:
$$ \chi^2 = \sum_{i=1}^{N}\left(\frac{I_i-I_i({\rm age,\zh,\afe})}{\sigma_{I_i}}\right)^2,$$
where $N$ is the number of Lick indices involved in the analysis, $I_i$ is an observed index,
$\sigma_{I_i}$ is the total uncertainty of the index, including rms error
of transformations to the Lick/IDS system, $I_i({\rm age,\zh,\afe})$ is
the theoretical index prediction.

A contour plot on Figure~B.1--6 shows isolines of three-dimensional $\chi^2$
solution space, sliced at the point of the global minimum. The contour $\chi^2$ levels and
the corresponding 67\%\, 95\%\, and 99\%\ confidence plots are shown.
See Section~3.1.1 for a discussion about the advantages and limitations of using of Lick indices and
the $\chi^2$ minimization technique for evolutionary parameters determination.
\begin{figure}[!t]
\label{GC7chi2}
\includegraphics[width=9.0cm]{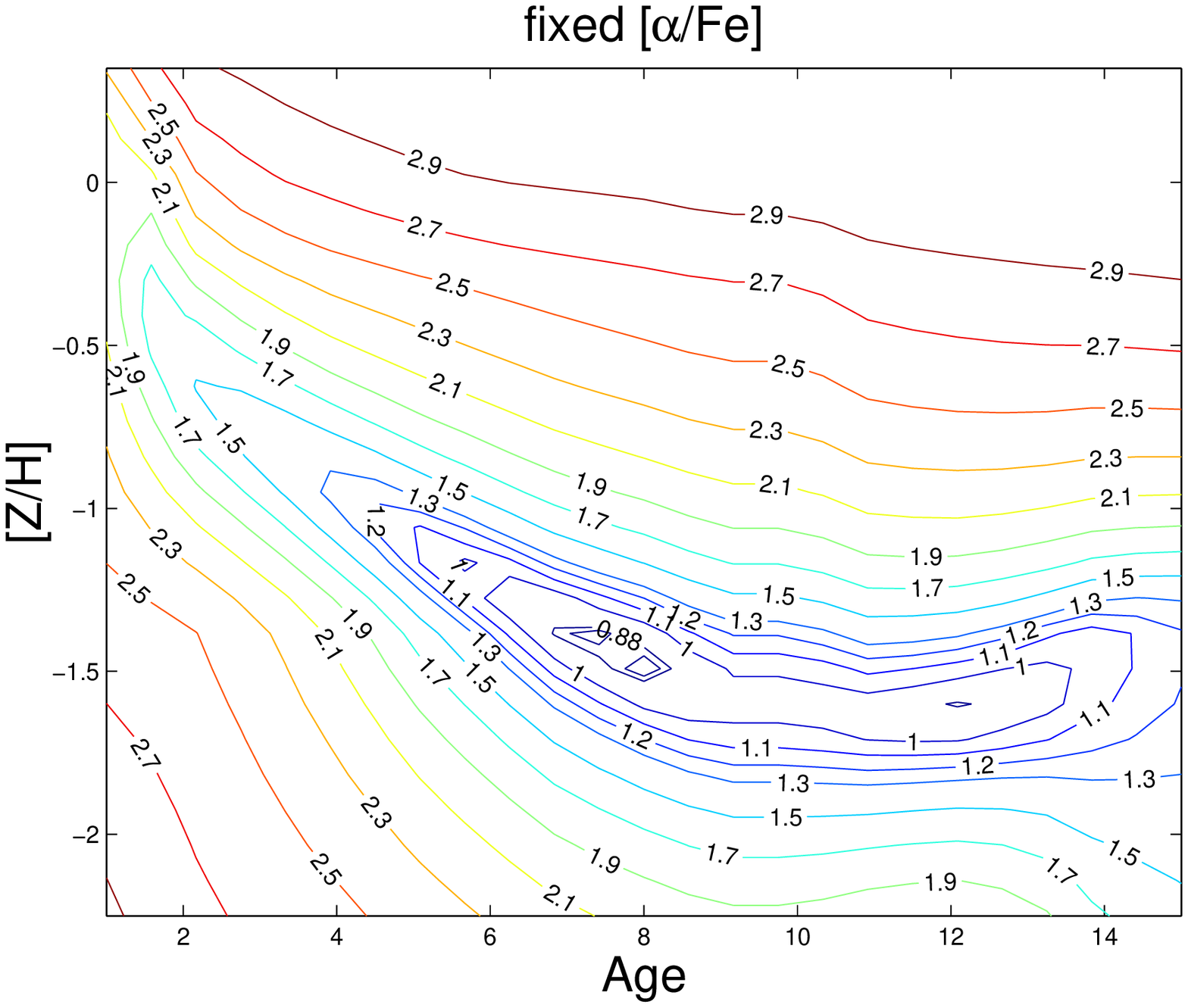}
\includegraphics[width=9.0cm]{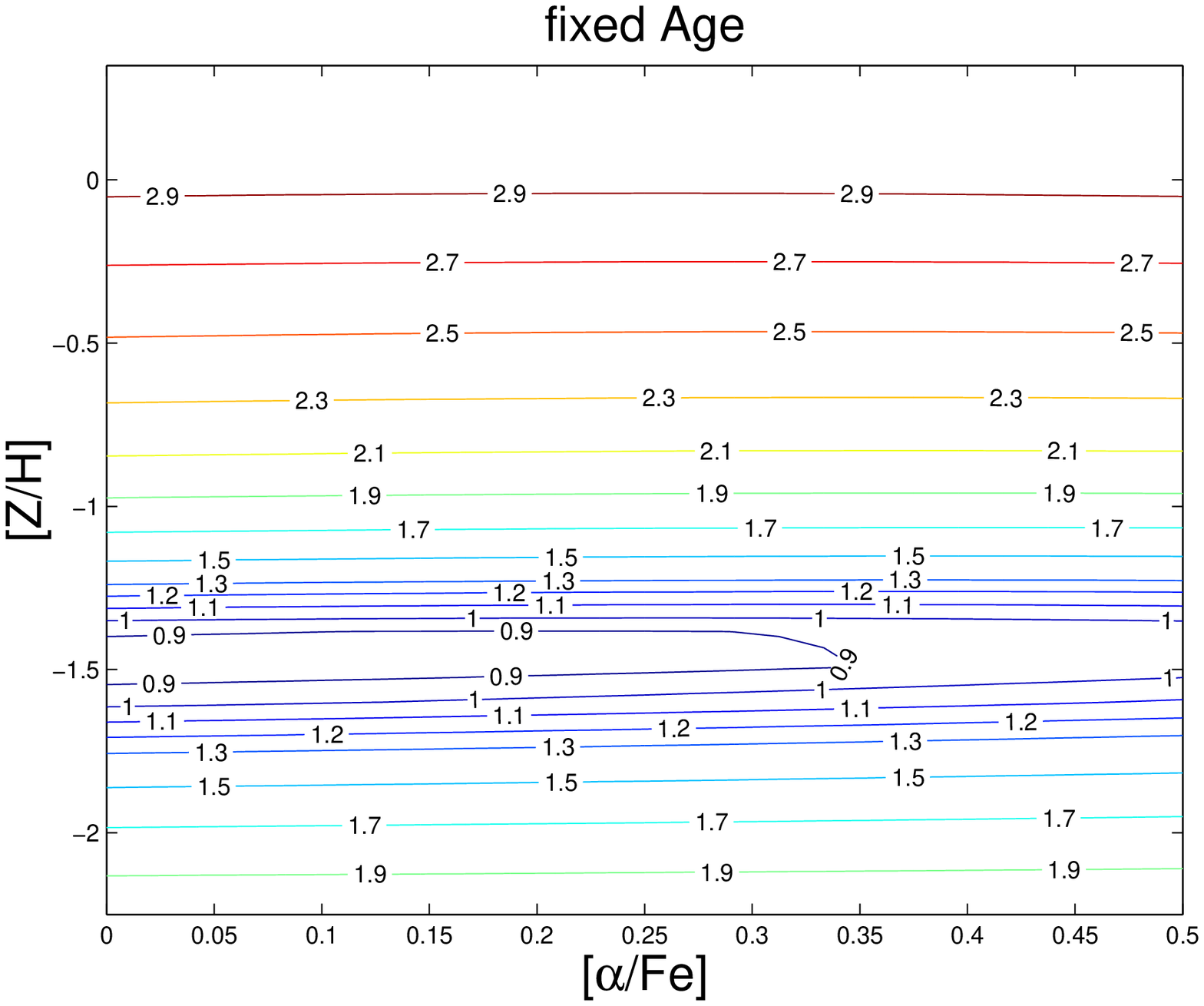}
\includegraphics[width=9.0cm]{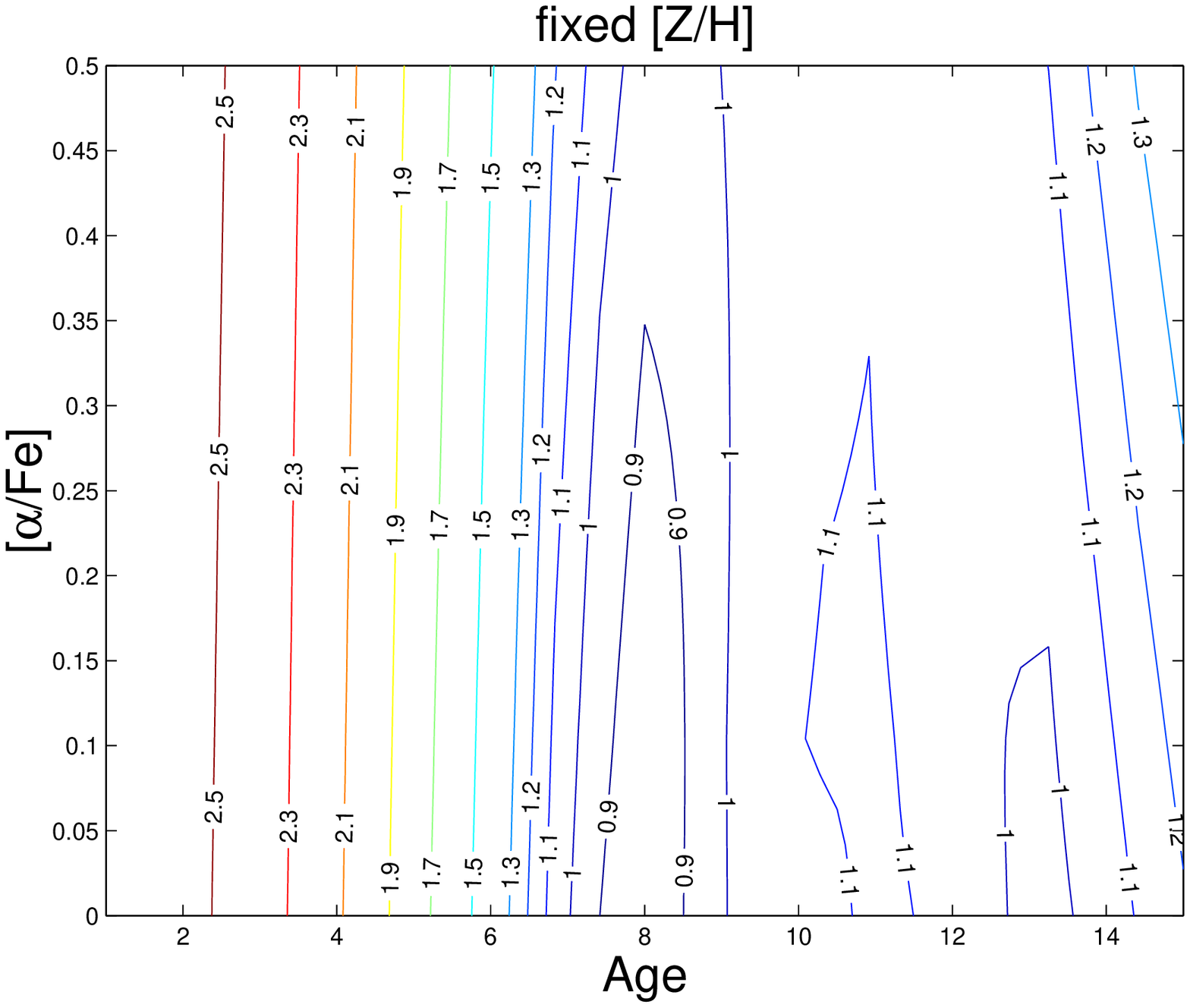}
\caption{{\bf GC7:} Age -- \zh, \afe\ -- \zh, and Age -- \afe\ slices of the three-dimensional $\chi^2$ space taken
at the point of the global minimum.}
\end{figure}
\begin{figure}
\includegraphics[width=9.0cm]{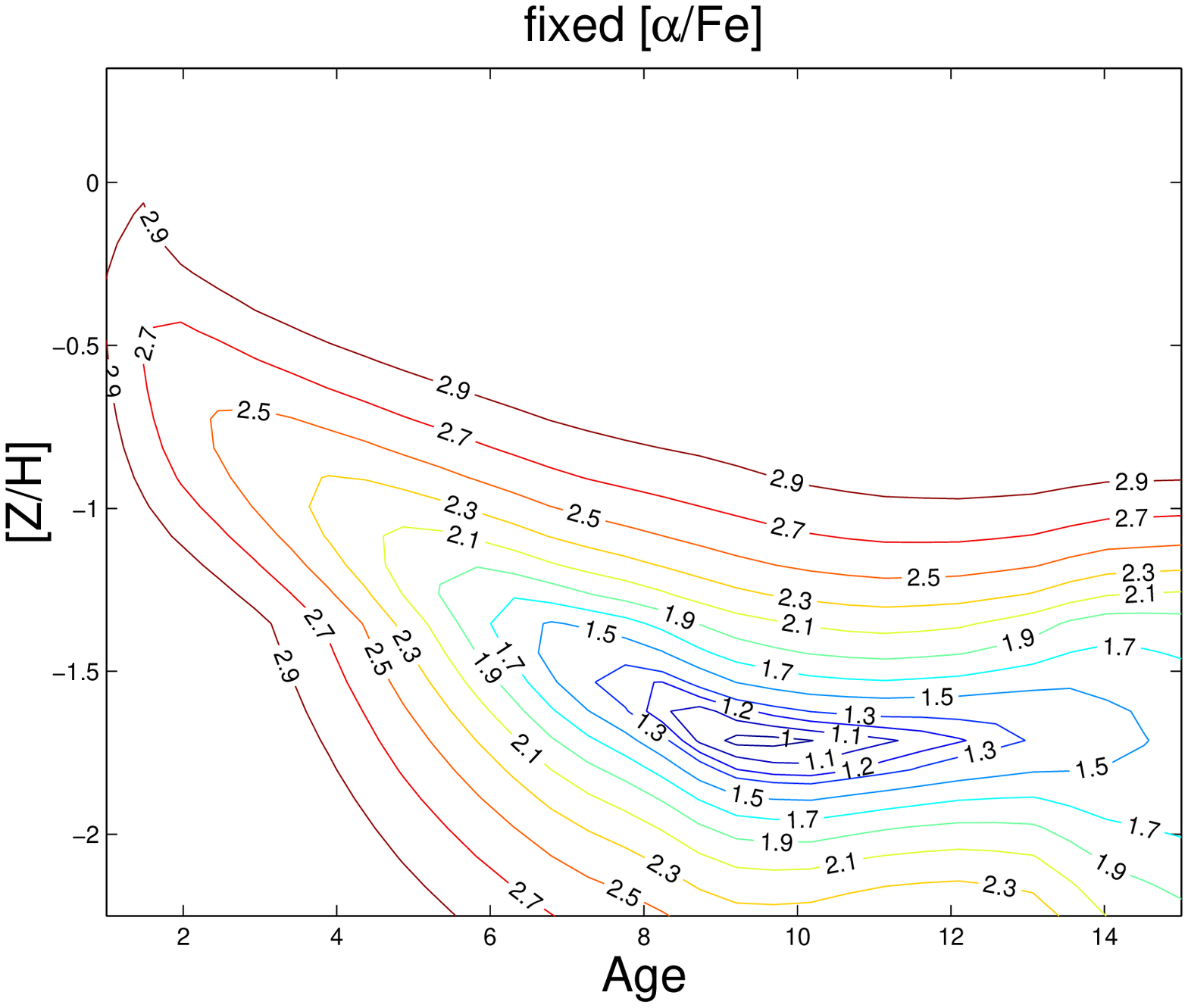}
\includegraphics[width=9.0cm]{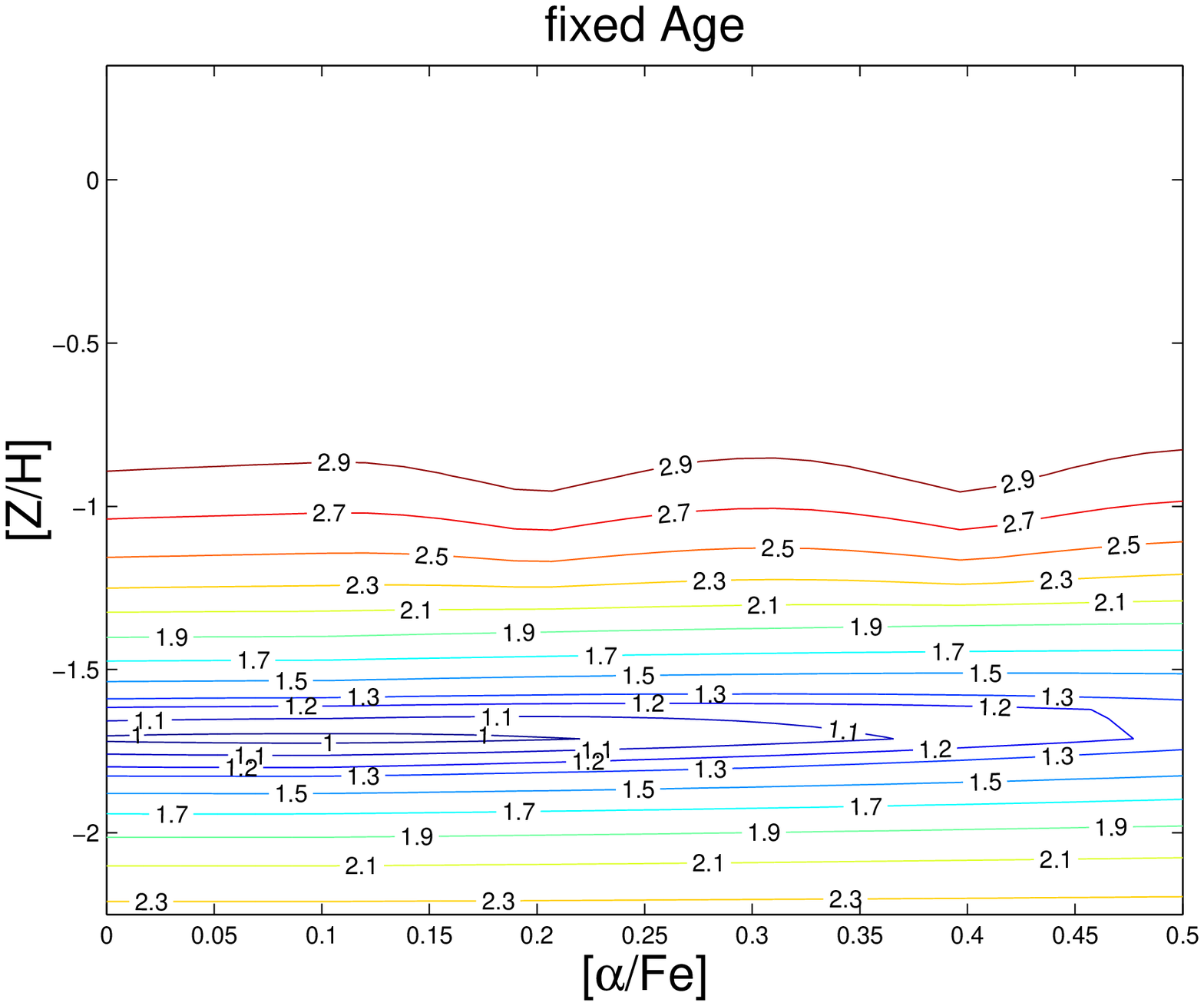}
\includegraphics[width=9.0cm]{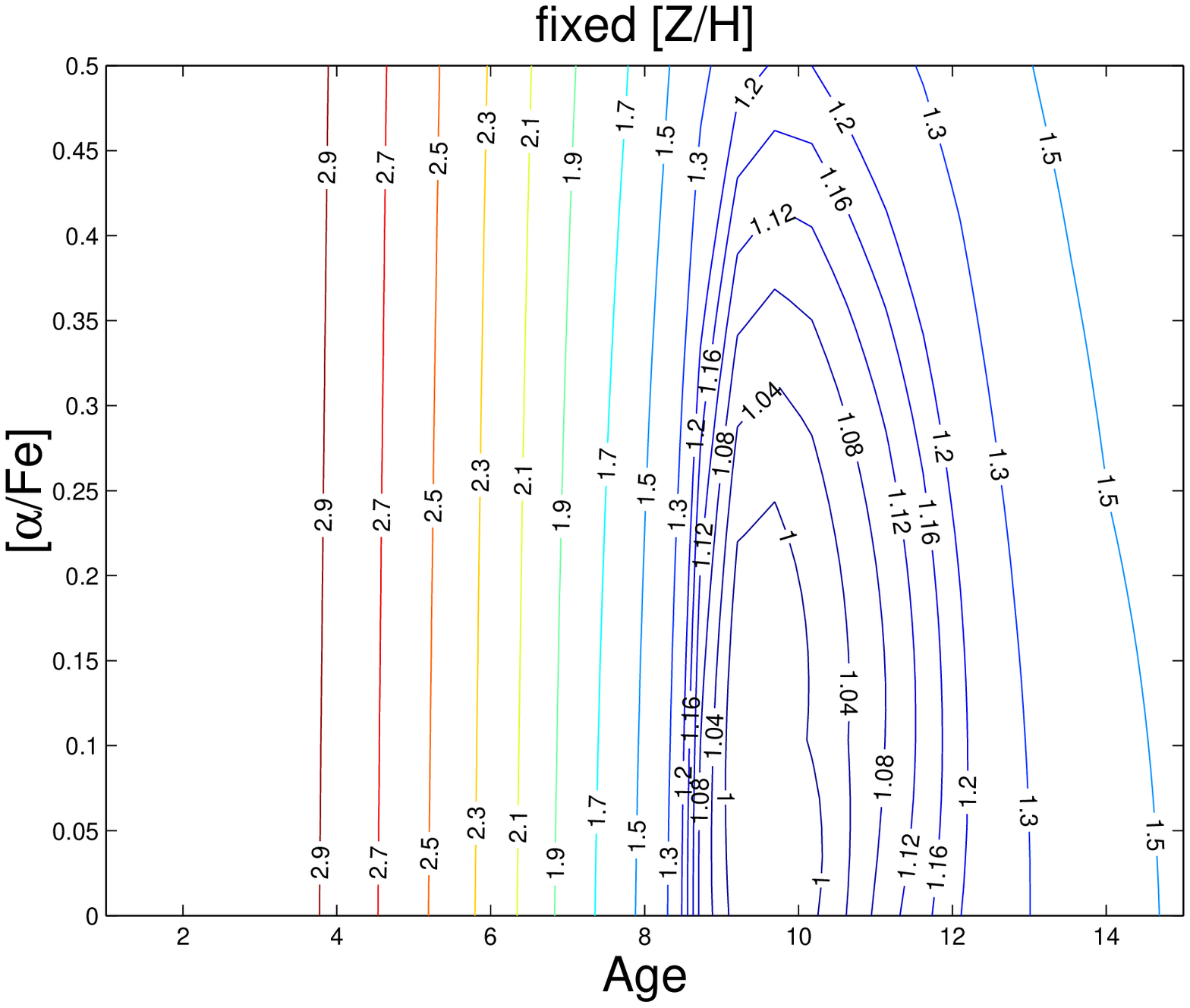}
\caption{Same as in Fig.~\ref{GC7chi2}, but for GC5.}
\end{figure}
\begin{figure}
\includegraphics[width=9.0cm]{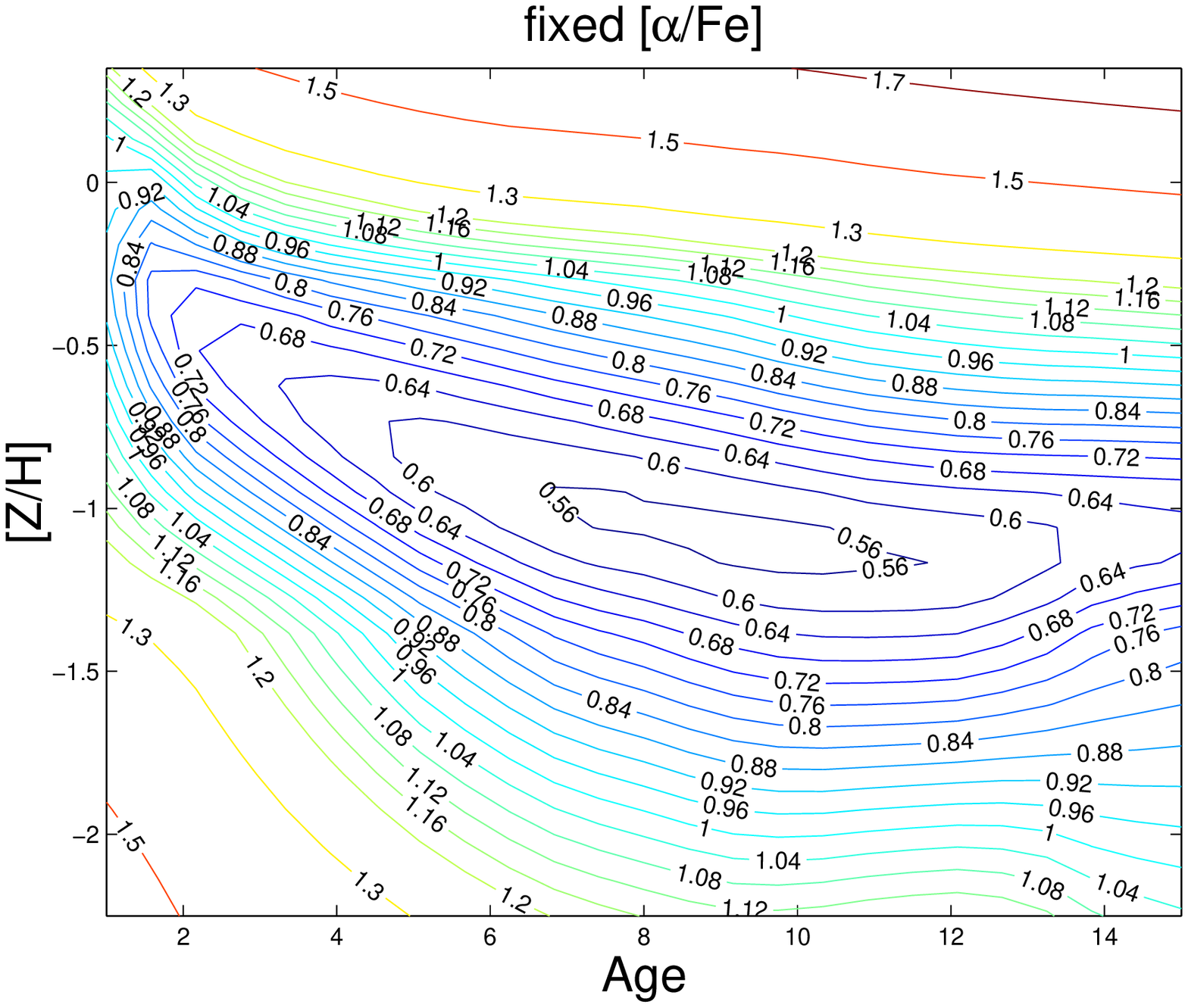}
\includegraphics[width=9.0cm]{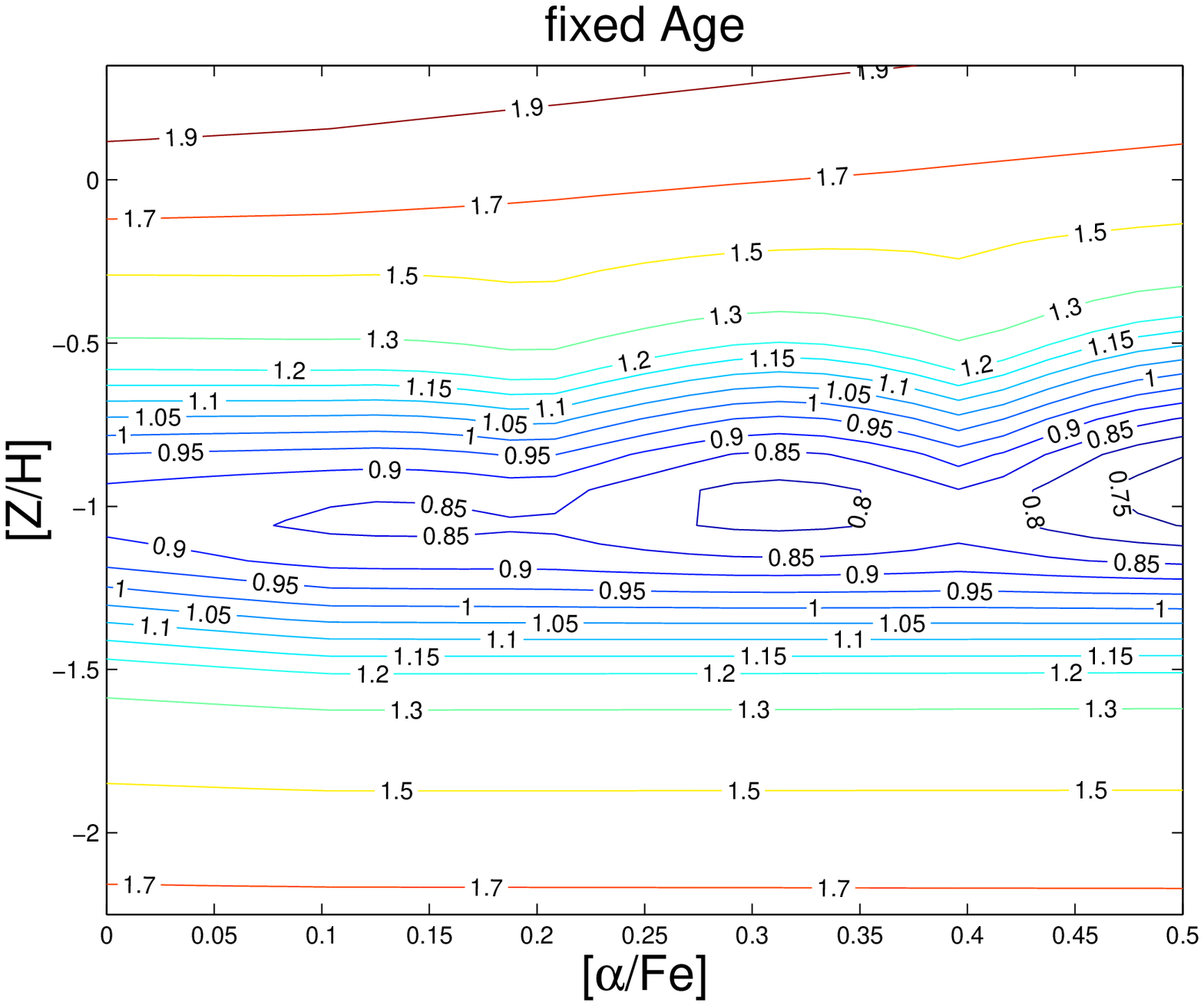}
\includegraphics[width=9.0cm]{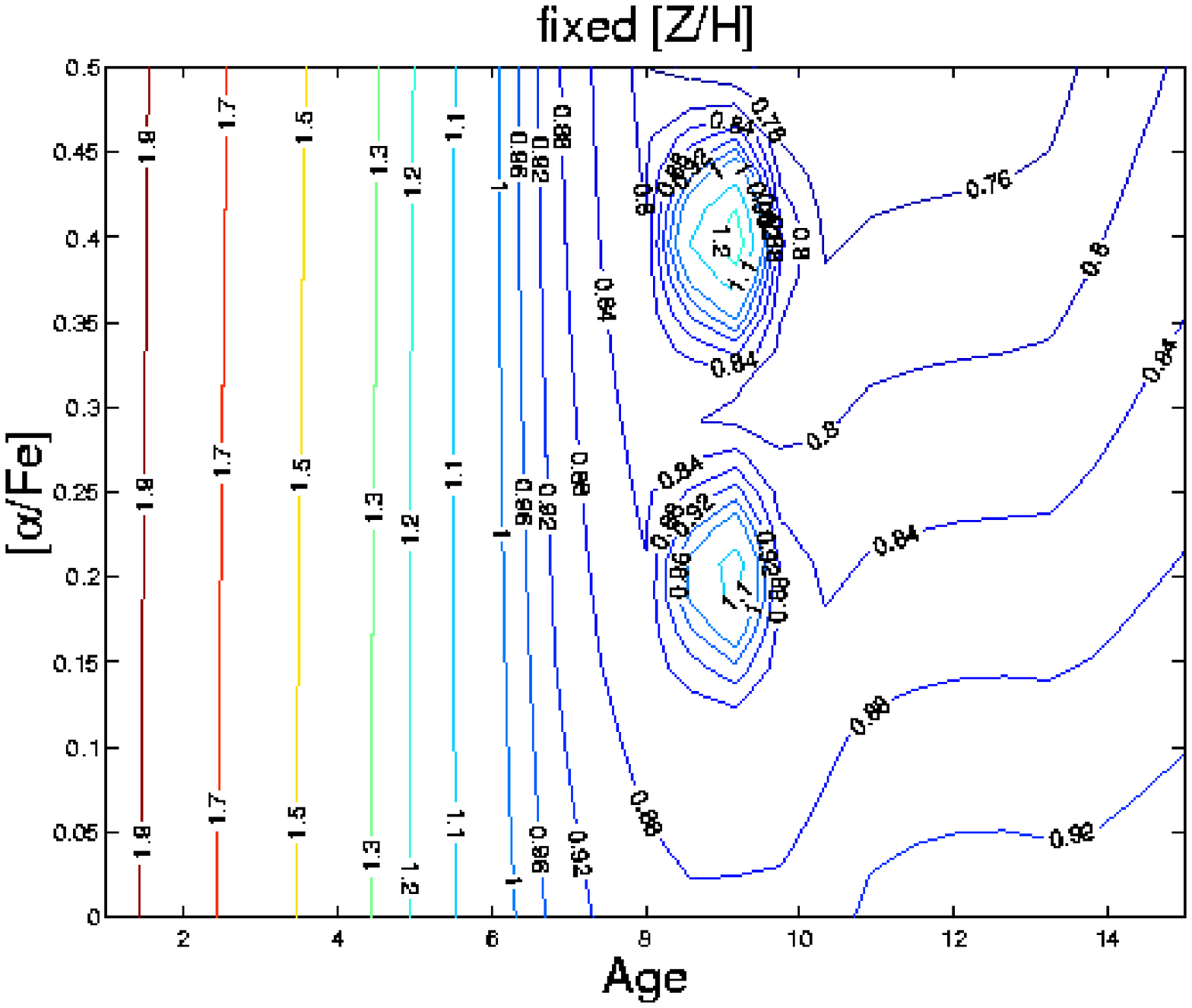}
\caption{Same as in Fig.~\ref{GC7chi2}, but for Hodge~II.}
\end{figure}
\begin{figure}
\includegraphics[width=9.0cm]{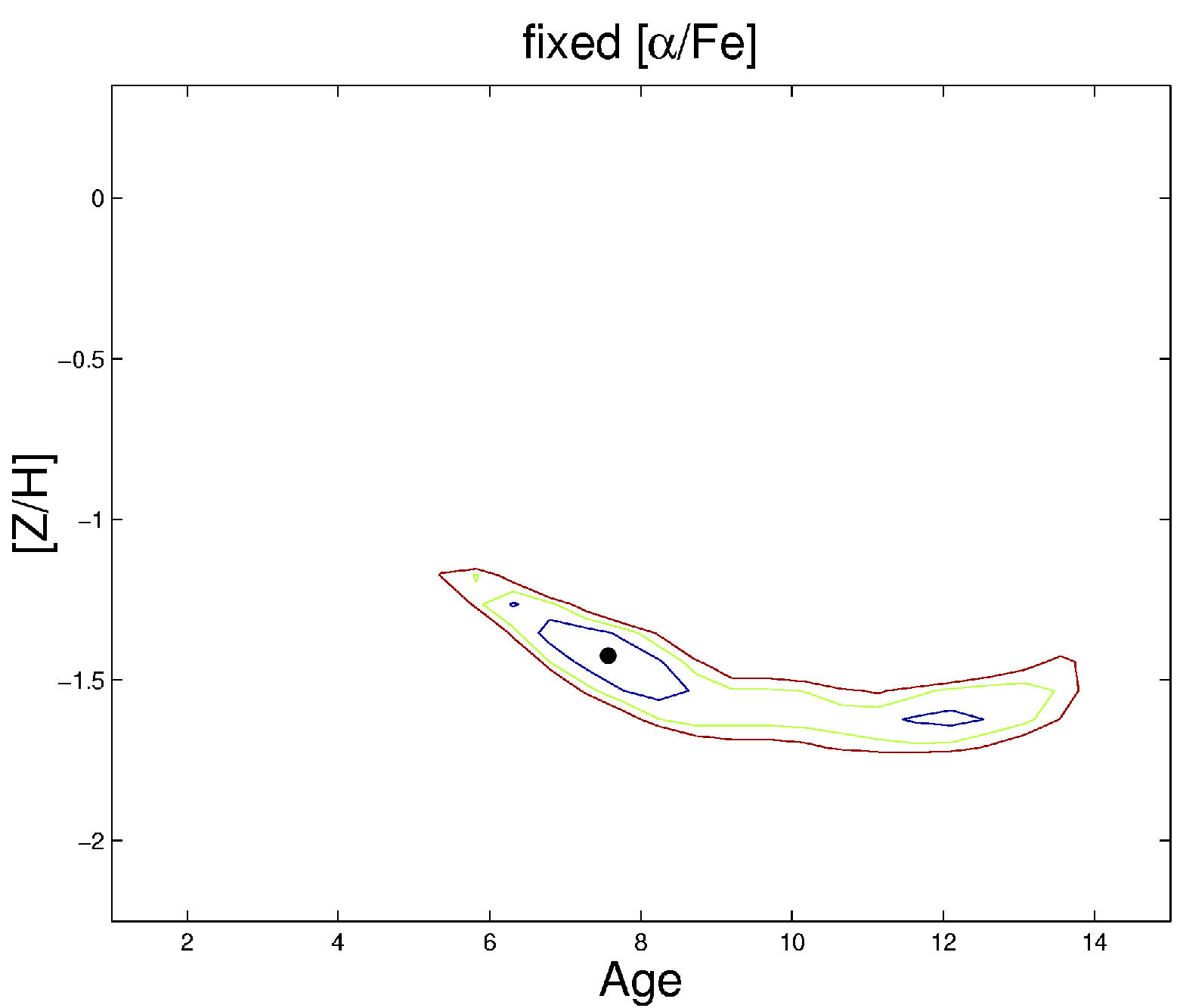}
\includegraphics[width=9.0cm]{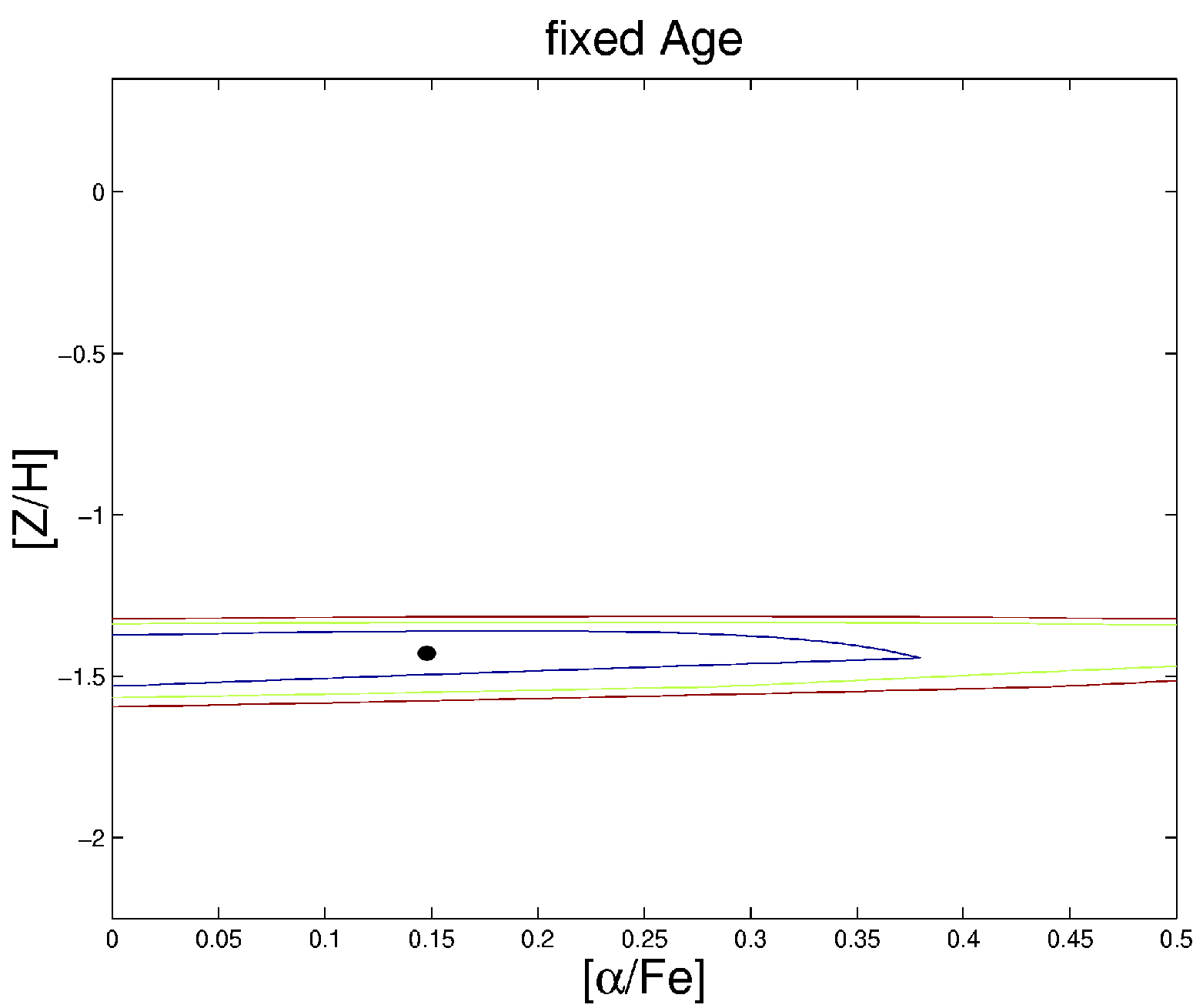}
\includegraphics[width=9.0cm]{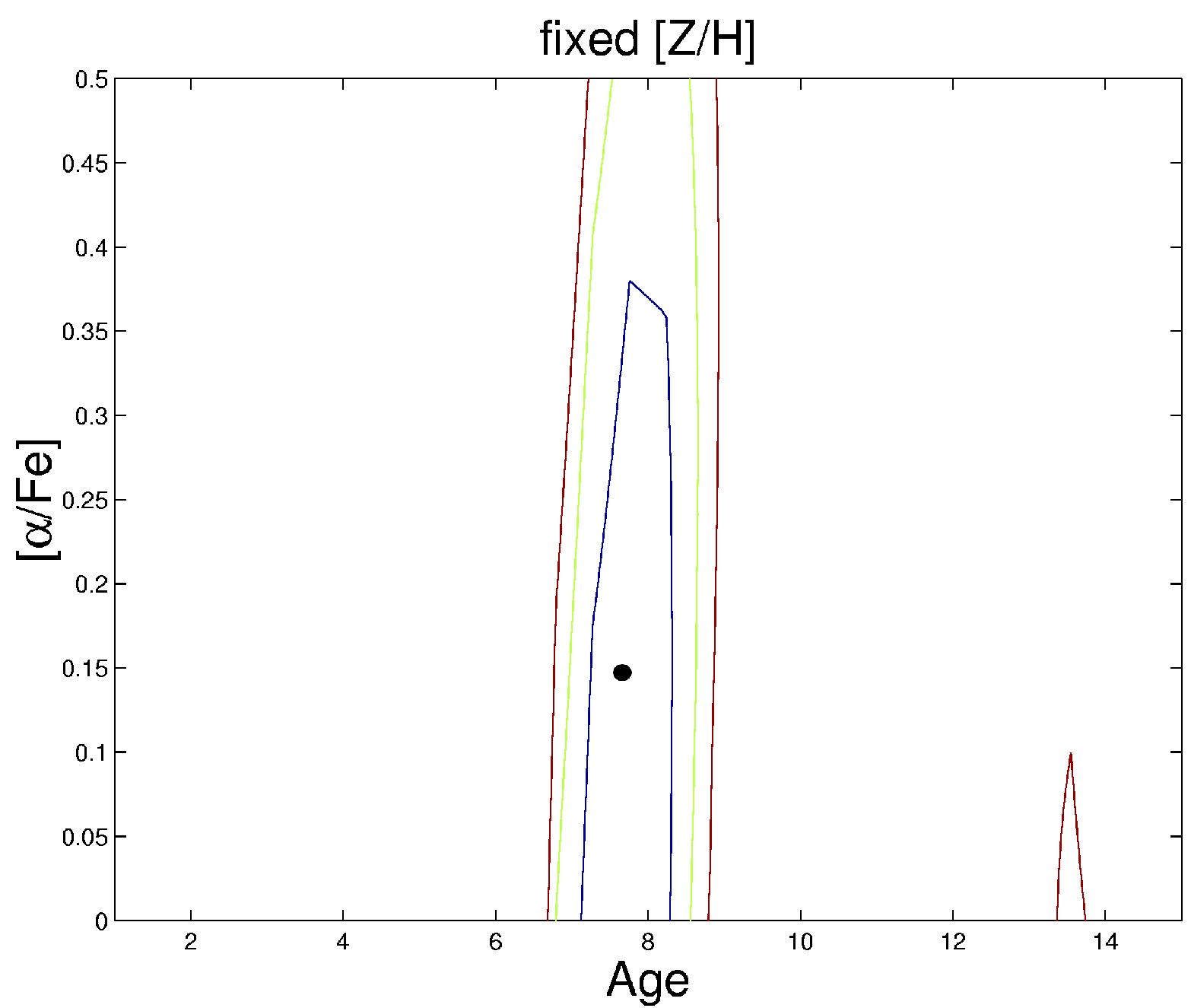}
\caption{{\bf GC7:} 67\%\, 95\%\, and 99\%\ confidence plots in the Age -- \zh, \afe\ -- \zh, and Age -- \afe\ planes.}
\end{figure}
\begin{figure}
\includegraphics[width=9.0cm]{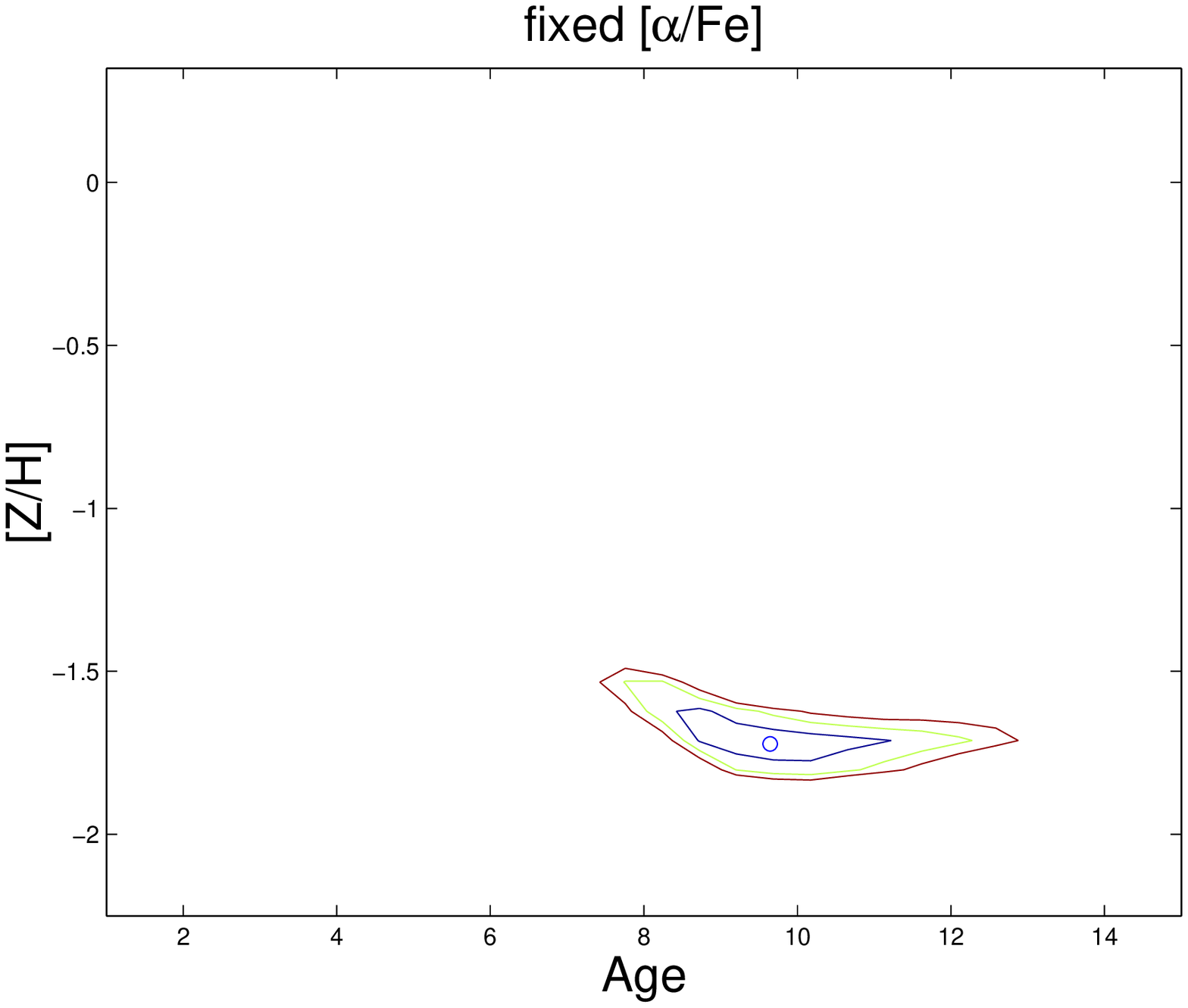}
\includegraphics[width=9.0cm]{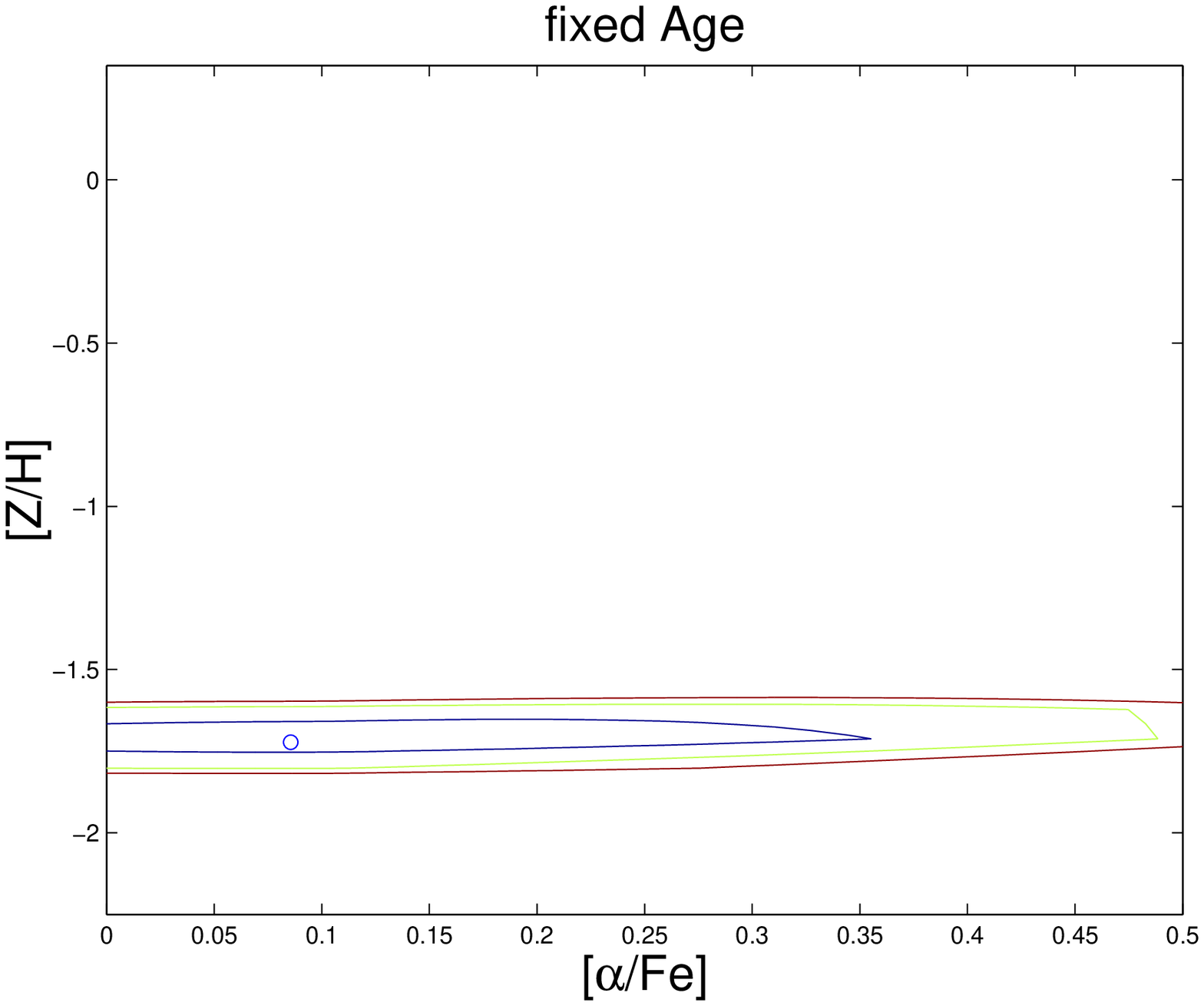}
\includegraphics[width=9.0cm]{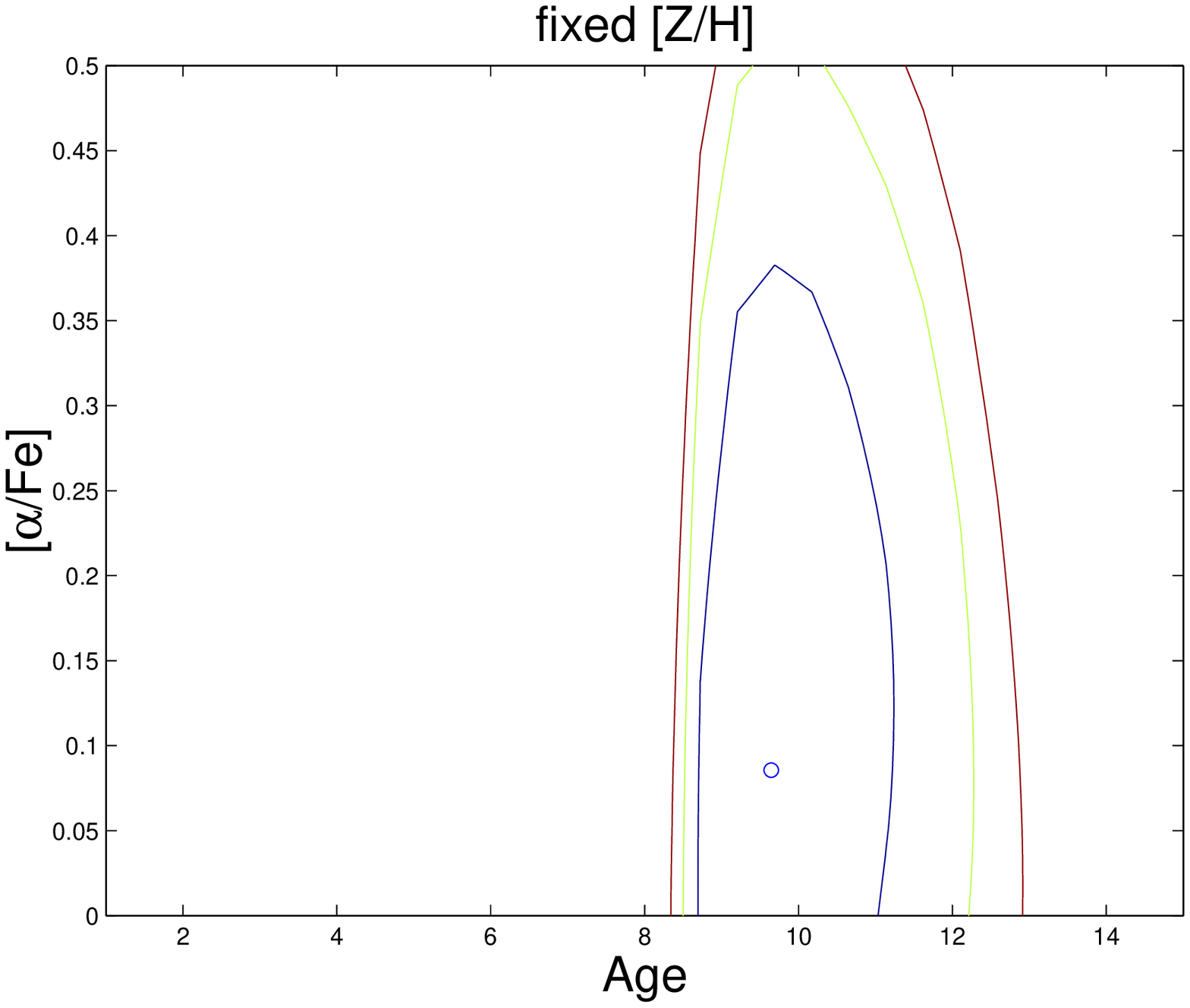}
\caption{Same as in Fig.~\ref{GC7chi2}, but for GC5.}
\end{figure}
\begin{figure}
\includegraphics[width=9.0cm]{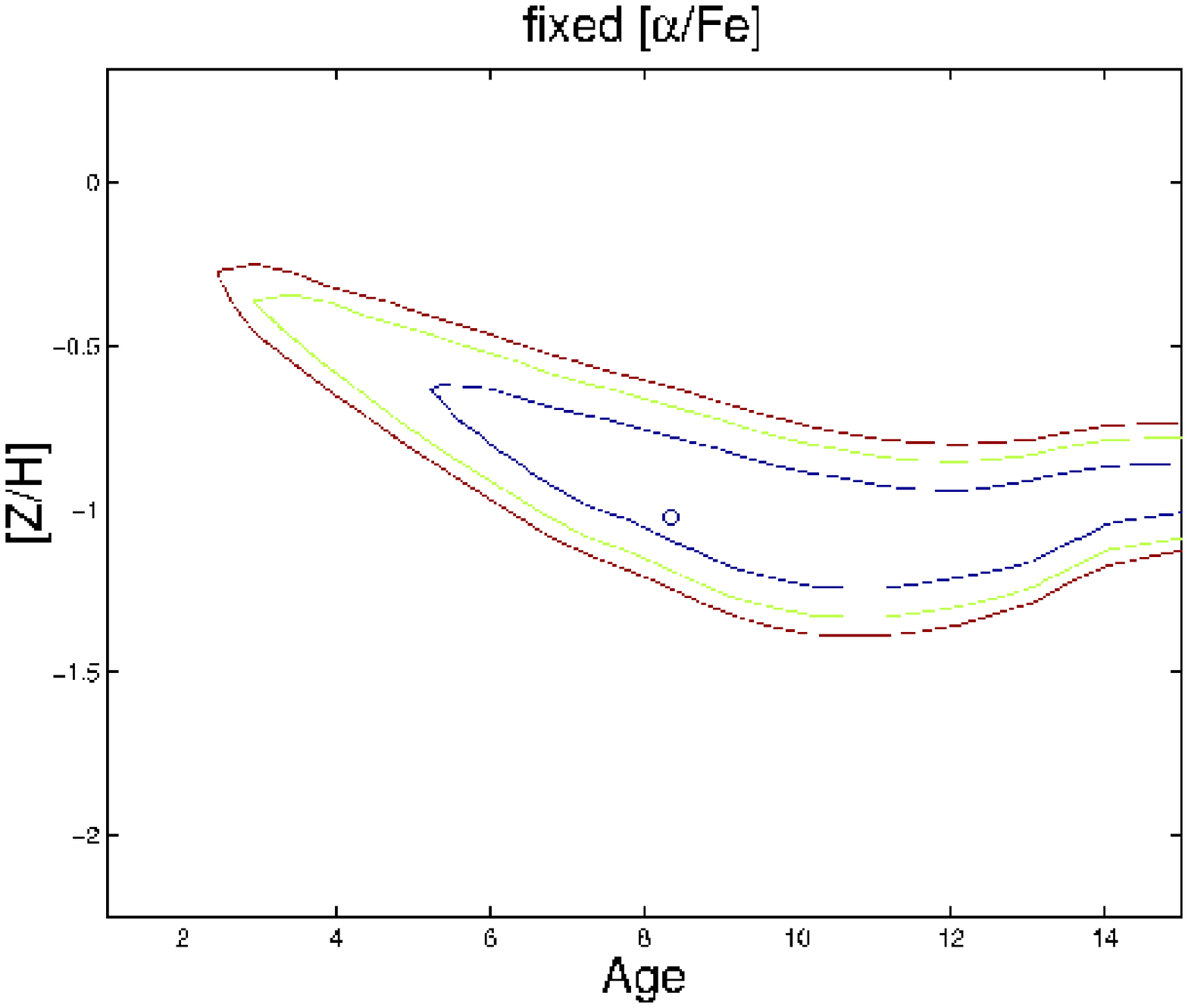}
\includegraphics[width=9.0cm]{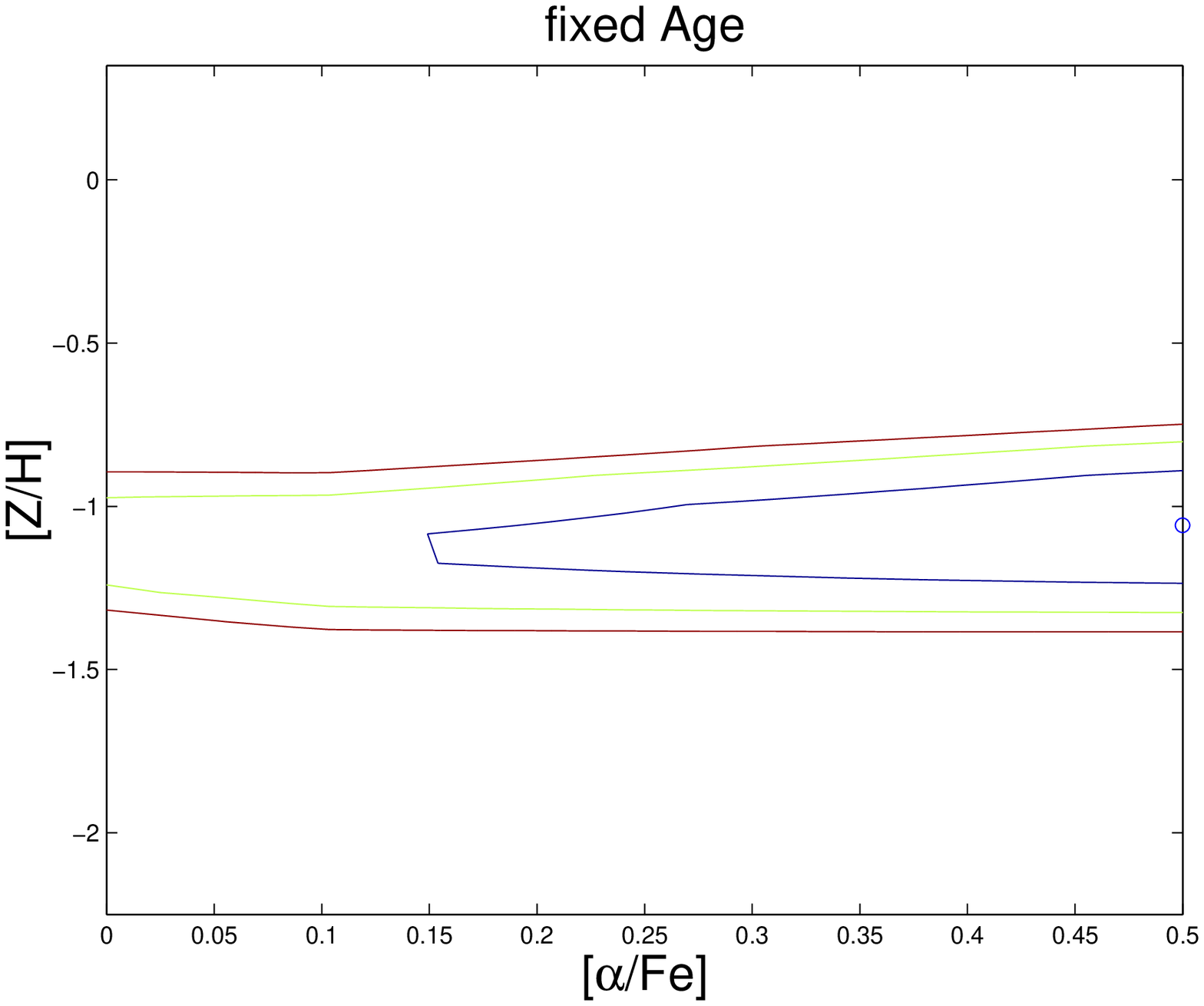}
\includegraphics[width=9.0cm]{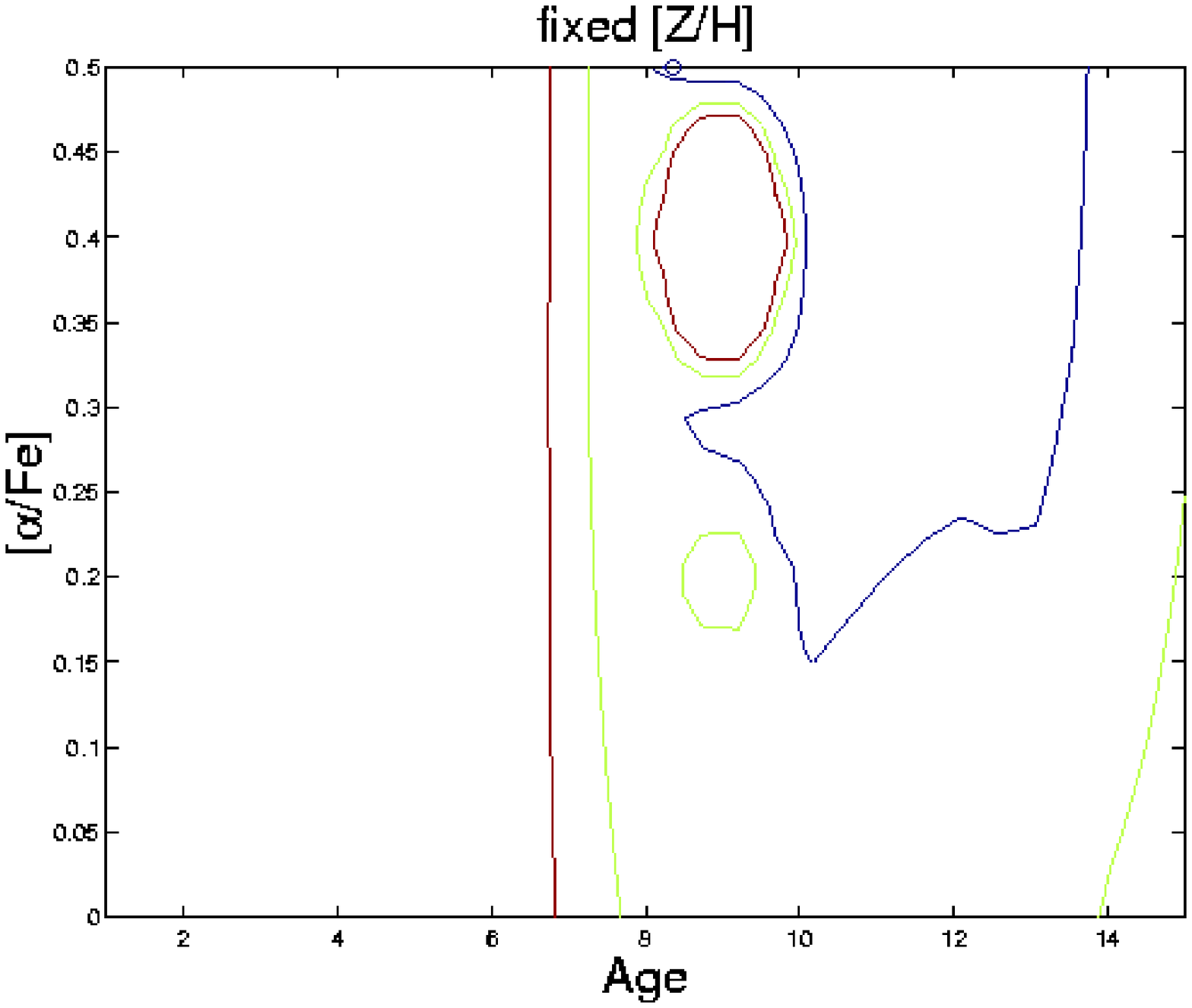}
\caption{Same as in Fig.~\ref{GC7chi2}, but for Hodge~II.}
\end{figure}

\section{Lick Index Measurements}
\begin{table*}[!b]
\caption{Globular cluster indices ($\lambda \le 4531$\AA\ ) (first line)
corrected for zeropoints of transformation to the standard Lick system
and errors determined from bootstrapping of the object spectrum (second line).
The S/N ratios per pixel in the spectra of the clusters  Hodge~II, GC5 and GC7,
measured at 5000 \AA\ are respectively $\sim$ 26, 30 and 99.}
\label{lickind1}
\begin{tabular}{llrrrrrrrrrrr} \\
\hline \hline
ID                 & H $\delta_{\rm A}$ & H $\gamma_{\rm A}$& H $\delta_{\rm F}$ & H $\gamma_{\rm F}$ & CN$_1$  & CN$_2$ & Ca4227 & G4300 & Fe4383 & Ca4455    \\
		   &(\AA)             & (\AA)          &  (\AA)              &     (\AA)             & (mag)    & (mag   & (\AA)  & (\AA) & (\AA)  & (\AA)     \\
 \hline
\noalign{\smallskip}
Hodge~II             &  2.92            &  0.46          &   2.48             &   0.99                & -0.118   &  -0.074 &   0.46 &  3.91 & -0.51  &   0.65   \\
 \hskip 50pt $\pm$   &  0.20            &  0.20          &   0.21             &   0.21                &  0.002   &   0.003 &   0.11 &  0.11 &  0.13  &   0.13   \\
GC5                 &  3.65            &  1.58          &   2.45             &   2.32                & -0.156   &  -0.114 &  -0.17 &  1.95 &  1.47  &   0.48   \\
 \hskip 50pt $\pm$   &  0.18            &  0.18          &   0.19             &   0.19                &  0.002   &   0.003 &   0.10 &  0.11 &  0.12  &   0.12   \\
GC7                 &  3.36            &  1.98          &   2.58             &   2.24                & -0.141   &  -0.076 &   0.17 &  1.66 &  0.01  &   0.56   \\
 \hskip 50pt $\pm$   &  0.04            &  0.04          &   0.04             &   0.04                &  0.001   &   0.001 &   0.02 &  0.02 &  0.03  &   0.03   \\
\hline  \hline
\end{tabular}
\end{table*}
\begin{table*}
\caption{Globular cluster indices ($\lambda \ge 4531$\AA\ ) (first line)
corrected for zeropoints of transformation to the standard Lick system
and errors determined from bootstrapping of the object spectrum (second line).}
\label{lickind2}
\begin{tabular}{llrrrrrrrrr} \\
\hline \hline
ID                 & Fe4531 & Fe4668 & H$\beta$ & Fe5015 & Mg1     & Mg2       & Mgb   & Fe5270 & Fe5335 & Fe5406 \\
		   & (\AA) & (\AA)  &  (\AA)   & (\AA)  & (mag)    & (mag)     & (\AA) & (\AA)  & (\AA)  & (\AA)  \\
 \hline
\noalign{\smallskip}
Hodge~II            & 3.32  &   1.61  &   2.94  &   2.21 &  0.031  &   0.040  &  0.67  &   1.56  &   0.81   &    1.48 \\
 \hskip 50pt $\pm$  & 0.14  &   0.16  &   0.16  &   0.17 &  0.005  &   0.005  &  0.18  &   0.18  &   0.18   &    0.18 \\
GC5                & 0.99  &   0.75  &   2.56  &   1.57 &  0.007  &   0.017  &  1.14  &   0.70  &   0.33   &    0.51 \\
 \hskip 50pt $\pm$  & 0.13  &   0.14  &   0.15  &   0.16 &  0.004  &   0.004  &  0.16  &   0.16  &   0.17   &    0.17 \\
GC7                & 0.72  &   0.82  &   2.19  &   1.61 &  0.016  &   0.030  &  0.62  &   0.50  &   0.65   &    0.75 \\
 \hskip 50pt $\pm$  & 0.03  &   0.03  &   0.03  &   0.03 &  0.001  &   0.001  &  0.04  &   0.04  &   0.04   &    0.04 \\
\hline  \hline
\end{tabular}
\end{table*}

\end{document}